\documentclass[twocolumn]{aastex63}

\usepackage{gensymb}
\usepackage[utf8]{inputenc}
\usepackage{multirow}
\usepackage{amsmath}
\usepackage{xcolor}
\usepackage[normalem]{ulem}

\graphicspath{{./}{figures/}}
\providecommand{\sorthelp}[1]{} % Required for Planck bibfile

\begin{document}

\title{The Atacama Cosmology Telescope: Microwave Intensity and Polarization Maps\\ of the Galactic Center}

\author[0000-0002-1697-3080]{Yilun~Guan}
\affiliation{Department of Physics and Astronomy, University of Pittsburgh, Pittsburgh, PA, USA 15260}
\email{yilun.guan@pitt.edu}
\author[0000-0002-7633-3376]{Susan~E.~Clark}
\affiliation{Institute for Advanced Study, 1 Einstein Drive, Princeton, NJ 08540, USA}
\author[0000-0001-7449-4638]{Brandon~S.~Hensley}
\affiliation{Department of Astrophysical Sciences, Peyton Hall, Princeton University, Princeton, NJ, USA 08544}
\author[0000-0001-9731-3617]{Patricio~A.~Gallardo}
\affiliation{Department of Physics, Cornell University, Ithaca, NY 14853, USA}
\author[0000-0002-4478-7111]{Sigurd~Naess}
\affiliation{Center for Computational Astrophysics, Flatiron Institute, New York, NY, USA 10010}
\author{Cody~J.~Duell}
\affiliation{Department of Physics, Cornell University, Ithaca, NY 14853, USA}

% ----------------------------------------------
% Opt-in authors
% ----------------------------------------------

\author[0000-0002-1035-1854]{Simone~Aiola}
\affiliation{Center for Computational Astrophysics, Flatiron Institute, New York, NY, USA 10010}
\author[0000-0002-2287-1603]{Zachary~Atkins}
\affiliation{Joseph Henry Laboratories of Physics, Jadwin Hall, Princeton University, Princeton, NJ, USA 08544}
\author[0000-0003-0837-0068]{Erminia~Calabrese}
\affiliation{School of Physics and Astronomy, Cardiff University, The Parade, Cardiff, CF24 3AA, UK}
\author[0000-0002-9113-7058]{Steve~K.~Choi}
\affiliation{Department of Physics, Cornell University, Ithaca, NY 14853, USA}
\affiliation{Department of Astronomy, Cornell University, Ithaca, NY 14853, USA}
\author[0000-0002-6151-6292]{Nicholas~F.~Cothard} 
\affiliation{Department of Applied and Engineering Physics, Cornell University, Ithaca, NY 14853, USA}
\author[0000-0002-3169-9761]{Mark~Devlin}
\affiliation{Department of Physics and Astronomy, University of Pennsylvania, 209 South 33rd Street, Philadelphia, PA, USA 19104}
\author[0000-0003-2856-2382]{Adriaan~J.~Duivenvoorden}
\affiliation{Joseph Henry Laboratories of Physics, Jadwin Hall, Princeton University, Princeton, NJ, USA 08544}
\author[0000-0002-7450-2586]{Jo~Dunkley}
\affiliation{Joseph Henry Laboratories of Physics, Jadwin Hall, Princeton University, Princeton, NJ, USA 08544}
\affiliation{Department of Astrophysical Sciences, Peyton Hall, Princeton University, Princeton, NJ, USA 08544}
\author{Rolando~D\"unner}
\affiliation{Instituto de Astrof\'isica and Centro de Astro-Ingenier\'ia, Facultad de F\'isica, Pontificia Universidad Cat\'olica de Chile, Av. Vicu\~na Mackenna 4860, 7820436, Macul, Santiago, Chile}
\author[0000-0003-4992-7854]{Simone~Ferraro}
\affiliation{Lawrence Berkeley National Laboratory, One Cyclotron Road, Berkeley, CA 94720, USA}
\affiliation{Berkeley Center for Cosmological Physics, UC Berkeley, CA 94720, USA}
\author[0000-0002-2408-9201]{Matthew~Hasselfield}
\affiliation{Center for Computational Astrophysics, Flatiron Institute, New York, NY, USA 10010}
\author[0000-0002-8816-6800]{John~P.~Hughes}
\affiliation{Department of Physics and Astronomy, Rutgers, the State University of New Jersey, 136 Frelinghuysen Road, Piscataway, NJ 08854-8019, USA}
\author[0000-0003-0744-2808]{Brian~J.~Koopman}
\affiliation{Department of Physics, Yale University, New Haven, CT 06520, USA}
\author[0000-0002-3734-331X]{Arthur~B.~Kosowsky}
\affiliation{Department of Physics and Astronomy, University of Pittsburgh, Pittsburgh, PA, USA 15260}
\author[0000-0001-6740-5350]{Mathew~S.~Madhavacheril}
\affiliation{Perimeter Institute for Theoretical Physics, 31 Caroline Street N, Waterloo ON N2L 2Y5 Canada}
\author{Jeff~McMahon}
\affiliation{Department of Astronomy, University of Chicago, Chicago, IL USA}
\affiliation{Department of Physics, University of Chicago, Chicago, IL 60637, USA}
\affiliation{Kavli Institute for Cosmological Physics, University of Chicago, 5640 S. Ellis Ave., Chicago, IL 60637, USA19}
\affiliation{Enrico Fermi Institute, University of Chicago, Chicago, IL 60637, USA}
\author[0000-0002-8307-5088]{Federico~Nati}
\affiliation{Department of Physics, University of Milano-Bicocca, Piazza della Scienza 3, 20126 Milano (MI), Italy}
\author[0000-0001-7125-3580]{Michael~D.~Niemack}
\affiliation{Department of Physics, Cornell University, Ithaca, NY 14853, USA}
\affiliation{Department of Astronomy, Cornell University, Ithaca, NY 14853, USA}
\affiliation{Kavli Institute at Cornell for Nanoscale Science, Cornell University, Ithaca, NY 14853, USA}
\author[0000-0002-9828-3525]{Lyman~A.~Page}
\affiliation{Joseph Henry Laboratories of Physics, Jadwin Hall, Princeton University, Princeton, NJ, USA 08544}
\author{Maria~Salatino}
\affiliation{Physics Department, Stanford University, Stanford 94305 CA, USA}
\affiliation{Kavli Institute for Particle Astrophysics and Cosmology, Stanford 94305 CA, USA}
\author{Emmanuel~Schaan}
\affiliation{Lawrence Berkeley National Laboratory, One Cyclotron Road, Berkeley, CA 94720, USA}
\affiliation{Berkeley Center for Cosmological Physics, UC Berkeley, CA 94720, USA}
\author[0000-0002-9674-4527]{Neelima~Sehgal}
\affil{Physics and Astronomy Department, Stony Brook University, Stony Brook, NY 11794}
\author[0000-0002-8149-1352]{Crist\'obal Sif\'on}
\affiliation{Instituto de F\'isica, Pontificia Universidad Cat\'olica de Valpara\'iso, Casilla 4059, Valpara\'iso, Chile}
\author[0000-0002-7020-7301]{Suzanne~Staggs}
\affiliation{Joseph Henry Laboratories of Physics, Jadwin Hall, Princeton University, Princeton, NJ, USA 08544}
\author{Eve~M.~Vavagiakis}
\affiliation{Department of Physics, Cornell University, Ithaca, NY 14853, USA}
\author[0000-0002-7567-4451]{Edward~J.~Wollack}
\affiliation{NASA / Goddard Space Flight Center, Greenbelt, MD, 20771, USA}
\author[0000-0001-5112-2567]{Zhilei~Xu}
\affiliation{Department of Physics and Astronomy, University of Pennsylvania, 209 South 33rd Street, Philadelphia, PA 19104, USA}
\affiliation{MIT Kavli Institute, Massachusetts Institute of Technology, 77 Massachusetts Avenue, Cambridge, MA 02139, USA}

\begin{abstract}
  We present arcminute-resolution intensity and polarization maps of
  the Galactic center made with the Atacama Cosmology Telescope. The
  maps cover a 32 deg$^2$ field at 98, 150, and 224 GHz with
  $\vert l\vert\le4^\circ$, $\vert b\vert\le2^\circ$. We combine these data with Planck
  observations at similar frequencies to create coadded maps with
  increased sensitivity at large angular scales. With the coadded
  maps, we are able to resolve many known features of the Central
  Molecular Zone (CMZ) in both total intensity and polarization. We
  map the orientation of the plane-of-sky component of the Galactic
  magnetic field inferred from the polarization angle in the CMZ,
  finding significant changes in morphology in the three frequency
  bands as the underlying dominant emission mechanism changes from
  synchrotron to dust emission. Selected Galactic center sources,
  including Sgr~A*, the Brick molecular cloud (G0.253+0.016), the
  Mouse pulsar wind nebula (G359.23-0.82), and the Tornado supernova
  remnant candidate (G357.7-0.1), are examined in detail. These data
  illustrate the potential for leveraging ground-based cosmic
  microwave background polarization experiments for Galactic
  science.\vspace{1cm}
\end{abstract}


\section{Introduction}
\label{sec:intro}
Some of the most extreme interstellar environments in the Galaxy are found 
in the Galactic center \citep[e.g.,][]{Battersby:2020}. The inner $\sim 500$\,pc of
the Milky Way is home to the Central Molecular Zone (CMZ), the densest
concentration of molecular gas in the Galaxy, with a mean density of
$\sim 10^4$\,cm$^{-3}$ \citep{Gusten:1989, Ferriere:2007}. The surface
density of dense gas greatly exceeds that found in nearby star-forming
molecular clouds, 
\added{with the average gas surface density transitioning from $\sim 5 M_{\odot}$\,pc$^{-2}$ to several hundreds $M_{\odot}$\,pc$^{-2}$ as one reaches the inner 200\,pc of the Galaxy \citep[see][for a review]{Morris:1996}.}
Standard prescriptions predict that the CMZ should
be an extremely active site of star formation, and yet the observed
star formation rate is low; by some estimates, an order of magnitude
or more below predictions \citep[e.g.,][and references
therein]{Longmore:2013, Barnes:2017, Nguyen:2021}. 

The apparently inefficient star
formation in the CMZ makes this region an ideal testbed for star formation theories,
with many factors proposed to explain the observations. These include the
strong magnetic field in the Galactic center \citep{Crutcher:1996, Chuss:2003, Ferriere:2011}, the rate of mass inflow to the CMZ \citep{Sormani:2019}, the strength and
compressibility of turbulence in the CMZ \citep{Federrath:2017}, and the possibility that we
are observing a relatively quiescent period between episodic bursts of
star formation \citep{Kruijssen:2014, Krumholz:2015}.
Furthermore, the CMZ is in some respects a nearby analog of nuclear
rings in other galaxies, including high-redshift starbursts. The Galactic center is thus an opportunity for up-close study of the physics
relevant to the cosmic history of star formation
\citep{Kruijssen:2013, Ginsburg:2019}.

The magnetic field in the vicinity of the Galactic center has long
been studied with radio polarimetry \citep{Ferriere:2009,
  Morris:2015}. The so-called nonthermal radio filaments -- thin
strands of radio-frequency emission -- were some of the earliest
observations to shed light on the magnetic field structure toward the
Galactic center. The nonthermal radio filaments are, for the most
part, strikingly perpendicular to the Galactic plane, and the
intrinsic magnetic field inferred from the Faraday de-rotated
polarized synchrotron emission tends to lie parallel to the long axis
of these filaments \citep{Morris:1985, Yusef-Zadeh:1987b, Lang:1999}.

Polarized dust emission provides a complementary means of probing
the magnetic field structure in the CMZ. Interstellar dust grains emit
partially polarized thermal radiation because they are aspherical and
preferentially align their short axes parallel to the ambient magnetic
field \citep{Purcell:1975}. The polarization angle of the dust emission is thus a
line-of-sight (LOS) integrated probe of the plane-of-sky component of the
magnetic field orientation. Polarized dust emission has been measured
at high angular resolution in small regions toward a number of CMZ
molecular clouds \citep[e.g.,][]{Novak:2000, Novak:2003, Chuss:2003,
  Matthews:2009, Roche:2018}. Recently, the balloon-borne experiment
PILOT presented a 240\,$\mu$m map of the polarized dust emission over
the entire CMZ region at $2.2'$ resolution \citep{Mangilli:2019},
along with comparisons to the lower-resolution 353\,GHz polarization
data measured by the Planck satellite \citep{planck2014-XIX}.

The Atacama Cosmology Telescope (ACT) measures the polarized microwave
sky with higher angular resolution than the Planck satellite and
greater sensitivity on small scales. In this paper we present new
dedicated maps of the Galactic center in total intensity and linear
polarization in three ACT frequency bands. We combine the ACT data
with Planck data to augment the map sensitivity on larger angular
scales. The frequency coverage of the maps presented here probe a
range of physical emission mechanisms, enabling a comprehensive view
of the Galactic center environment. In polarization these maps probe
both polarized dust and synchrotron emission, and in total intensity
the maps additionally show features from free-free emission and
molecular line emission from transition frequencies that fall within
the ACT passbands. These data illustrate the potential of sensitive
cosmic microwave background (CMB) polarization experiments for
Galactic science.

We describe the ACT observations in Section~\ref{sec:observation} and
the mapmaking and Planck coadd procedures in
Section~\ref{sec:mapmaking}. In Sections~\ref{sec:imaps} and
\ref{sec:pmaps}, we present the maps in total intensity and
polarization, respectively, and discuss derived properties, including
emission mechanisms, magnetic field orientation, and polarization
fraction. In Section~\ref{sec:regions}, we identify notable Galactic
center objects and compare to observations at other frequencies. We
conclude in Section~\ref{sec:conclusion}.

\section{Observations}
\label{sec:observation}
ACT is a 6-meter off-axis Gregorian telescope located at an elevation
of 5190\,m on Cerro Toco in the Atacama Desert in Chile \citep{Fowler:2007,Thornton:2016}. ACT scans the
millimeter-wave sky with arcminute resolution, complementary to the full-sky
lower angular-resolution measurements from satellite missions such as
the Wilkinson Microwave Anisotropy Probe \citep[WMAP;][]{Bennett:2013} and Planck \citep{planck2013-p01}.

In 2019, the target ACT observing fields were expanded to include the
Galactic center region. Between 2019 June 6 and November 29, a total
duration of $\sim35$ hr of data were taken with three Advanced ACTPol
dichroic detector arrays PA4, PA5, and PA6
\citep{henderson/etal:2016,ho/2017,choi/etal:2018}, at three frequency
bands f090, f150, and f220 centered roughly at 98, 150, and 224\,GHz,
respectively. The beam full-width-half-maximum (FWHM) at each band is
$2.0'$, $1.4'$, and $1'$, respectively. The observation field extends
roughly from $-89\degree$ to $-97\degree$ in declination and
$-33\degree$ to $-25\degree$ in right ascension. This study focuses
specifically on a 32 deg$^2$ field near the CMZ with Galactic
longitude $\vert l\vert \le 4^\circ$ and Galactic latitude $\vert b\vert \le 2^\circ$.

In this paper we present the maps made using the nighttime
observations only, which constitute roughly two-thirds of the total
data collected. The daytime observations are affected by a
time-dependent beam deformation due to the heating from the Sun that
is challenging to correct for in detailed high-resolution maps, and
hence those data are excluded from this analysis. Correcting for this
beam deformation will be a subject of future study, and the daytime
observations may be included for future versions of these maps.

\section{Mapmaking}\label{sec:mapmaking}
\subsection{Mapmaking with ACT}
% ============================================================
% total intensity planck vs. coadd side by side
% ============================================================
\begin{figure*}[ht]
  \centering
  \makebox[\textwidth][c]{\includegraphics[width=1.15\textwidth]{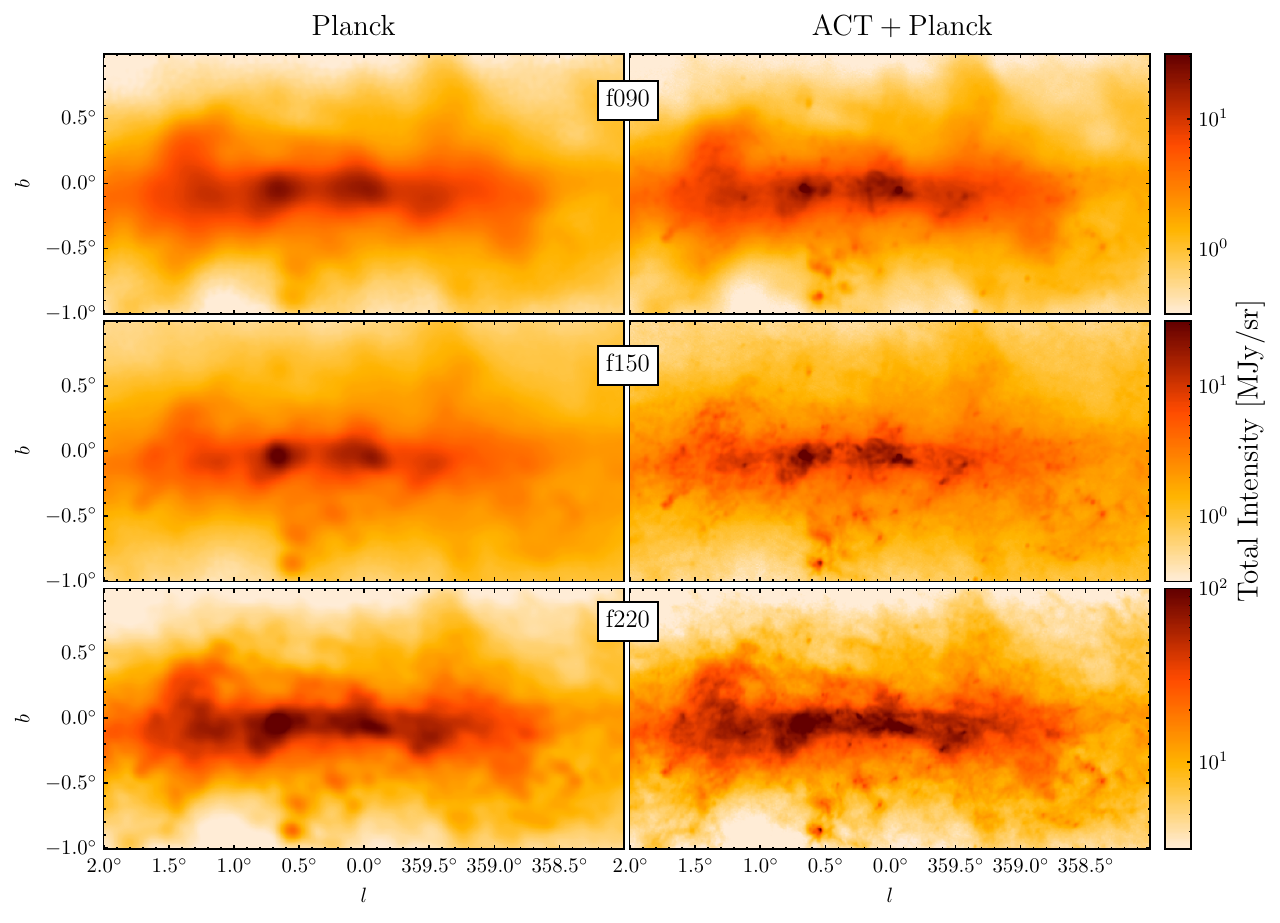}}
  \caption{Comparison between Planck-only maps (left column) and ACT+Planck
    coadded maps (right column) in total intensity. Rows from top to
    bottom correspond to f090, f150, and f220 respectively. Each map
    extends from $\vert l\vert \le 2^\circ$, $\vert b\vert \le 1^\circ$ and is plotted on a logarithmic color
    scale from 0.3--30 MJy\,sr$^{-1}$ for f090 and f150, and from
    from 3--100 MJy\,sr$^{-1}$ for f220. \added{See Figure~\ref{fig:t_maps_act} in Appendix~\ref{app:beam} for a corresponding plot with ACT-only maps.}} \label{fig:t_maps}
\end{figure*}
% ============================================================
% polarized intensity planck vs. coadd side by side
% ============================================================
\begin{figure*}[ht]
  \centering
  \makebox[\textwidth][c]{\includegraphics[width=1.1\textwidth]{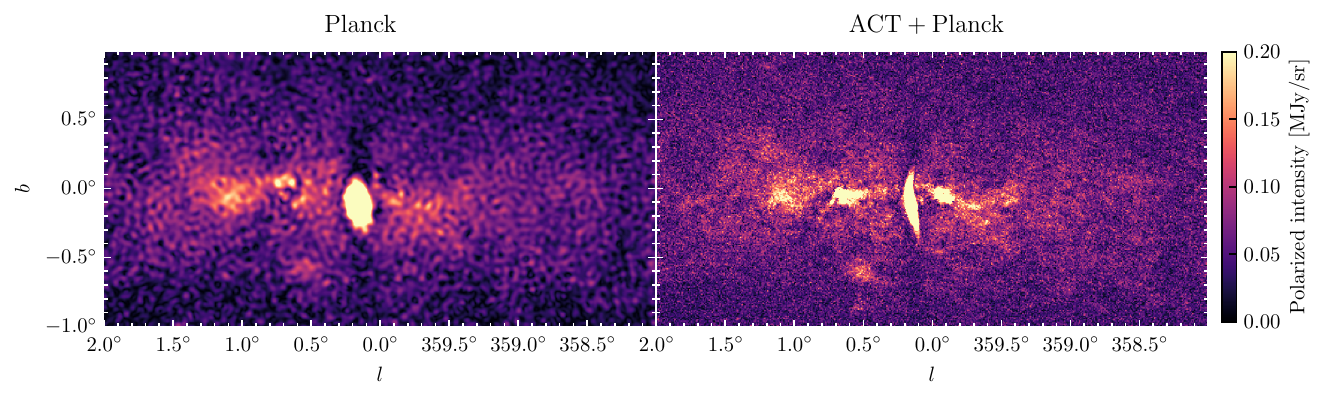}}
  \caption{A side-by-side comparison between Planck only (left) and the
    ACT+Planck coadded (right) for f150 in polarized intensity.}\label{fig:pol_sidebyside}
\end{figure*}
% ============================================================
% Q/U products in 3 bands
% ============================================================
\begin{figure*}[ht]
  \centering
  \makebox[\textwidth][c]{\includegraphics[width=1.1\textwidth]{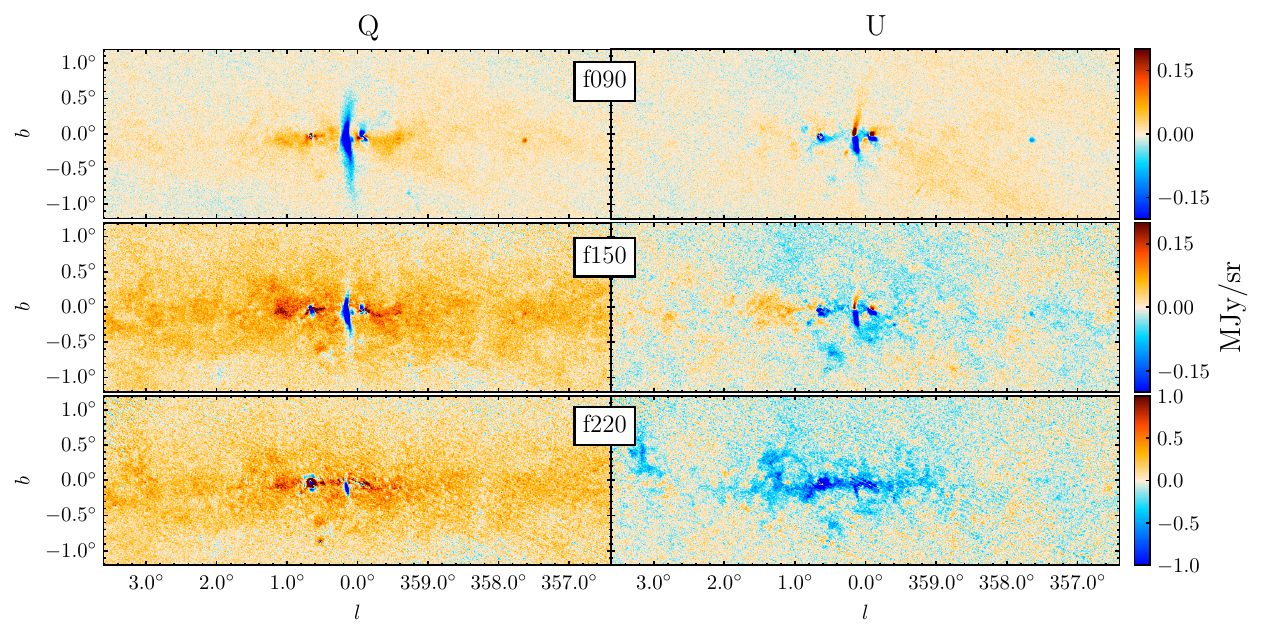}}
  \caption{Polarization maps in Stokes $Q$ (left column) and $U$
    (right column) in Galactic coordinates and using the IAU polarization
    convention. Top to bottom are the f090, f150, and f220 maps,
    respectively.}\label{fig:QU3bands}
\end{figure*}

The instrument records observations in the form of time-ordered data
(TOD) in units of $\sim$ 10 minutes. We largely follow the mapmaking pipeline
as described in \cite{Aiola:2020} with a few key differences, as we briefly summarize below.

First, we cut bad samples affected by glitches in each TOD. To prevent
bright sources in the Galactic center region from being mistaken for glitches, we
mask sources brighter than 5~mK with a radius of $3'$ prior to applying the glitch finder. \added{The 5~mK flux limit is chosen such that it is low enough to prevent bright sources from being mistaken as glitches, but high enough to ensure only a tiny fraction of sky is masked.} We also note that this mask is only applied when identifying glitches and not used during mapmaking. 
Timestreams with outlying statistical properties in terms of noise
levels and optical responsiveness are then flagged and removed from
the analysis. We further split the dataset into two independent
subsets for each frequency band and detector array respectively,
resulting in 12 datasets in total. We then obtain the sky maps for
each dataset by solving the mapping equation,
\begin{equation}
\label{eq:maping equation}
  d = Pm + n,
\end{equation}
for a set of Stokes parameters ($I$, $Q$, $U$), where $d$ is the
pre-processed time-streamed data, $P$ is the pointing matrix, $m$ is
the output map of interest, and $n$ is the noise model. This equation
yields a maximum-likelihood solution for $m$ by inverting
\begin{equation}
\label{eq:mapmaking}
  (P^TN^{-1}P)m=P^TN^{-1}d,
\end{equation}
where $N$ is the detector-detector noise covariance. 

There are two notable differences between the pipeline used in this
study and that presented in \cite{Aiola:2020}. First, we have adopted
a new calibration method that improves gain stability and reduces
biases from thermal contamination as compared to the method in ACT Data Release (DR) 4 
\citep{Aiola:2020} (see Appendix~\ref{app:calibration} for more details). The second
difference relates to the handling of point sources and extended hot
regions that are common in the Galactic center region but uncommon in
CMB fields. Directly applying the mapmaking pipeline in ACT DR4
leads to stripes around the bright sources caused by model errors
as explained in \cite{Naess:2019}. This happens for two reasons: (1) A pixelated 
map does not capture the sub-pixel behavior of the sky. These residuals 
are proportional to the gradient of the signal across a pixel and are 
often fractionally small. However, if the sky is sufficiently bright,
such as in the brightest parts of the Galactic center, they can still end
up being large in absolute terms. Since the map $m$ in 
Equation~\ref{eq:maping equation} cannot capture these 
residuals, the model forces them to be interpreted as part of the noise $n$. 
(2) The correlated noise model used in the mapmaker induces a nonlocal
response to the sub-pixel noise, leading to biases on the scale of the 
noise correlation length. To avoid this problem, we first identify the regions
that source the strongest model errors, namely, the brightest parts of the
Galaxy, and then eliminate model errors in these pixels by allocating an
extra degree of freedom for each sample that hit them, as described 
in \cite{Naess:2019}.

A caveat concerning these maps is that the bright parts of the Galaxy
were not masked when building the noise model $N$. The noise model
estimator assumes that the time-ordered data is noise dominated
($d \approx n$), and uses this to measure the noise covariance directly from
$d$. This breaks down when the telescope scans across the Galactic
center, resulting in an overestimate of the noise amplitude especially
on smaller scales. This has two consequences: (1) The data are
weighted sub-optimally in Equation~(\ref{eq:mapmaking}), resulting in
slightly higher noise. Since the maps are strongly signal dominated,
this can be ignored. (2) Because the noise model is contaminated by
the same signal it is applied to, there is a small loss in signal
power in the maps; pixels where noise happens to add constructively to
the signal have more power in $d$ than in pixels where they partially
cancel. Since we use inverse-variance weighting, the latter are
up-weighted compared to the former. The size of this effect is limited
because the problematically bright regions make up a small fraction of
the total samples used to build $N$. We have not measured the precise
size of this effect, but estimate it to be $\lesssim 1\%$ based on experience
with other high-signal-to-noise (S/N) regions, and hence we expect it
to have negligible impact on the interpretation of the maps in this
paper. This deficiency will be rectified in the upcoming ACT DR6 maps.

A final known issue requiring mitigation is
temperature-to-polarization (T-to-P) leakage. ACT typically scans a
given region of the sky both during its rising and setting. As the
Galactic center region is at relatively low declination, rising scans
and setting scans are poorly cross-linked (for more information on ACT
scan strategy see \citet{Stevens:2018}). Furthermore, the ACT beam is
known to leak T-to-P at the percent level. This beam leakage effect
averages down effectively in the nominal CMB maps, which are well
cross-linked, but in the Galactic center region the T-to-P leakage is
apparent at a $\sim 1$--$2\%$ level that contaminates the polarization
maps in the bright Galactic plane. To reduce the contamination from
beam leakage, we build a 2D leakage beam model for each dataset using
observations of Uranus made in the same observation year (2019), and
de-project the expected T-to-P leakage from the polarization maps in
each dataset (see Appendix~\ref{app:beam} for more details).

Following these methods, we produced two-way split maps of the Galactic center 
region at $0.5'$ resolution in Plate Carreé (CAR) projection for each
frequency band (f090, f150, f220) and detector array (PA4, PA5, PA6)
resulting in a total of 12 maps.

\subsection{Coadd with Planck}
% ============================================================
% multi-freq plot in I and P
% ============================================================
\begin{figure*}[ht]
  \centering
  \makebox[\textwidth][c]{\includegraphics[width=1.1\textwidth]{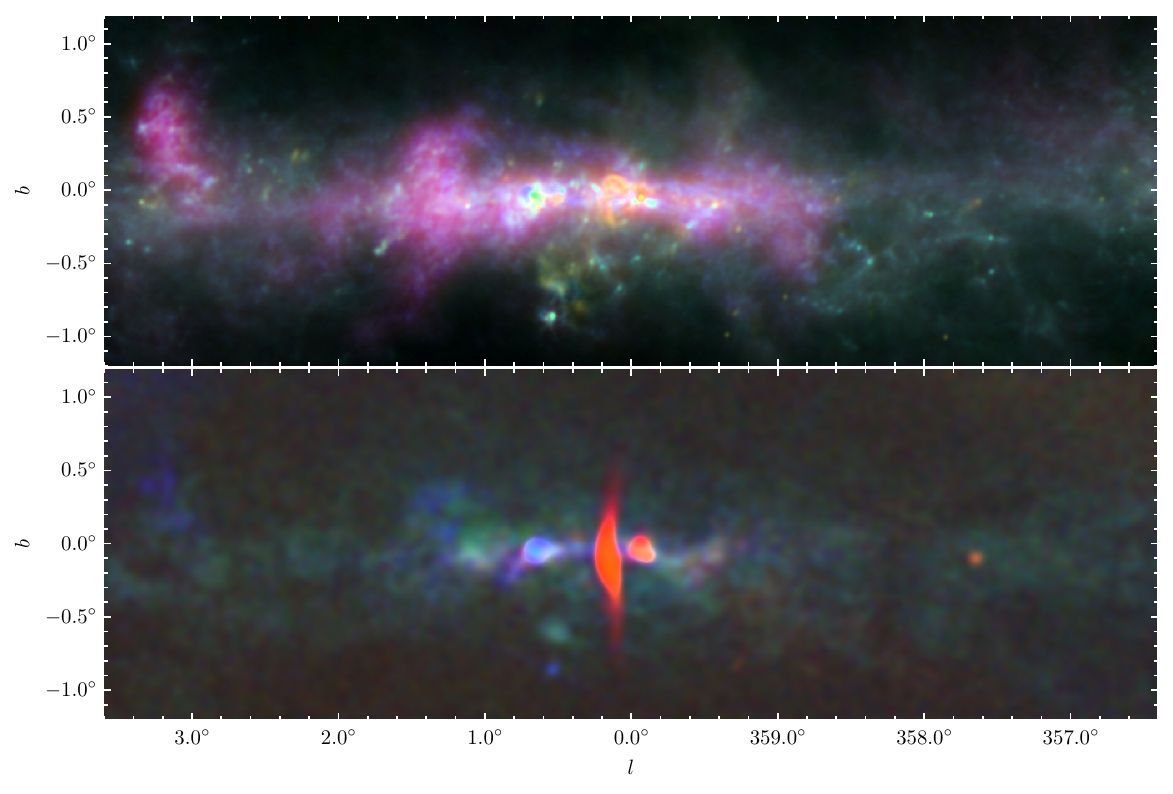}}
  \caption{Multifrequency view of the Galactic center region in both
    total intensity (upper panel) and polarized intensity (lower
    panel). Red, green, and blue correspond to f090, f150, and f220,
    respectively. In the upper panel, the maps are scaled
    logarithmically from 0.2 to 2~MJy\,sr$^{-1}$ for f090, from 0.214
    to 2.14~MJy\,sr$^{-1}$ for f150, and from 1.15 to
    10.15~MJy\,sr$^{-1}$ for f220. The polarization maps shown in the
    lower panel are first smoothed with a Gaussian kernel
    (FWHM$=3.5'$) and then scaled linearly from 0 to
    1~MJy\,sr$^{-1}$ for f090, to 1.79~MJy\,sr$^{-1}$ for
    f150, and to 8.2~MJy\,sr$^{-1}$ for f220.}\label{fig:multifreq}
\end{figure*}

The large angular scales in the ACT maps are affected by atmospheric
noise contamination and complicated co-variances at large scales. These modes can be recovered, however, by
coadding ACT maps with maps from Planck, which
dominate the S/N at large scales $\ell \lesssim 1000$. In
particular, we have used a similar algorithm as presented in ACT DR5
\citep{Naess:2020}, in combination with the Planck
High Frequency Instrument (HFI) maps processed through the NPIPE
pipeline \citep{planck2020-LVII}, which are two-way split maps
featuring improved noise level and systematic control as compared to
the previous Planck data releases.
\begin{figure*}[ht]
  \centering
  \includegraphics[width=\textwidth]{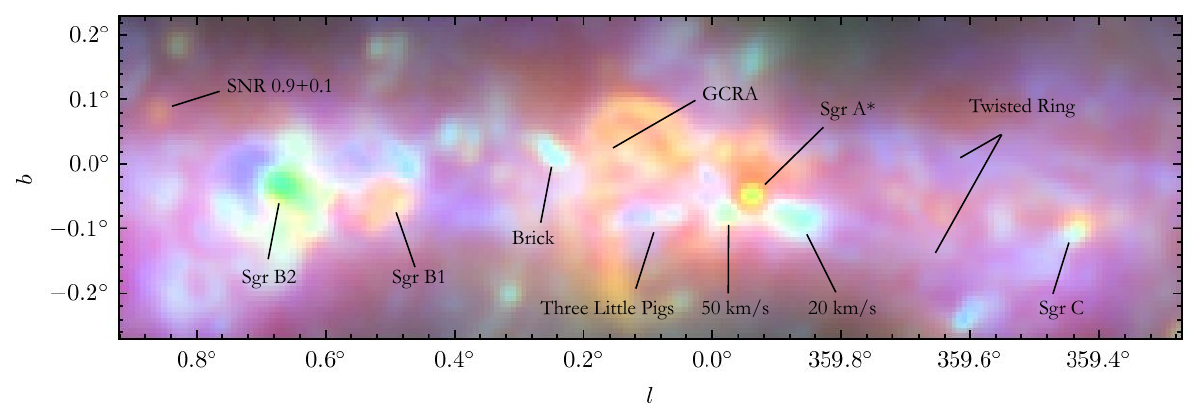}
  \includegraphics[width=\textwidth]{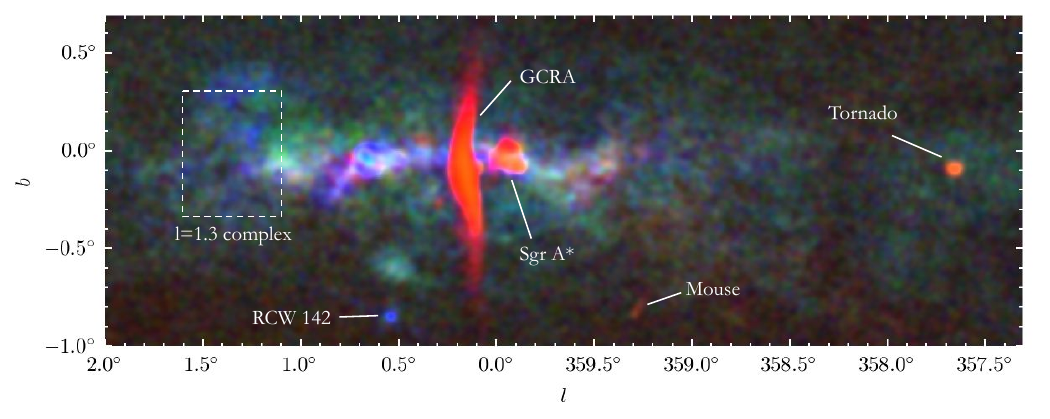}
  \caption{\textbf{Upper panel}: Known radio sources found in the
    Galactic center region. The background image shows a zoomed-in
    view of the multifrequency 3-color image presented in the upper
    panel of Figure~\ref{fig:multifreq}. \textbf{Lower panel}:
    Annotations of selected radio and dusty sources in the
    multifrequency polarized intensity image (presented in the lower
    panel of Figure~\ref{fig:multifreq}). Note that we used a
    smoothing with FWHM=$2'$ to make objects more
    visible.}\label{fig:annot}
\end{figure*}

As the coadding algorithm is presented in detail in \cite{Naess:2020},
we only briefly summarize the steps and note differences here. First,
we re-project the Planck maps and noise models from
HEALPix\footnote{\url{http://healpix.sf.net}} \citep{Gorski:2005}
projection with $N_{\rm side}=2048$ into the ACT Galactic center
observation footprint in CAR projection with $0.5'$ pixelization using
bi-cubic interpolation. We have used the same passbands as in
\citet{Naess:2020} and similarly matched the Planck 100\,GHz maps with
ACT f090 maps, 143\,GHz with ACT f150, and 217\,GHz with ACT f220.
This process assumes that the ACT and Planck passbands are equivalent.
We note that this introduces additional scale dependence to the
effective band centers \citep{Naess:2020}. This is expected to have
negligible impact on the results presented here but is relevant for
component-separation analysis, which will be the subject of follow-up
work.

We then solve for the maximum-likelihood coadded maps using a
block-diagonal equation
\begin{equation}
  \label{eq:coadd}
  \begin{pmatrix}
    m_0\\m_1\\\vdots
  \end{pmatrix} =
  \begin{pmatrix}
    B_0\\B_1\\\vdots
  \end{pmatrix}B_{\rm out}^{-1}m + n,
\end{equation}
where $m_i$ refers to each individual map, $B_i$ refers to its
corresponding beam transfer function, and $m$ refers to the final coadded
map with a desired beam $B_{\rm out}$, which is the ACT beam in this
case. $n$ refers to the map noise, which is assumed to be Gaussian and
block diagonal across individual maps, i.e., individual maps have
independent noise realizations. Of the noise models presented in
\citet{Naess:2020}, we have adopted the constant correlation noise model,
though the choice makes little difference in practice as we are
considering only a small patch of sky with close to uniform noise levels. We
invert Equation~(\ref{eq:coadd}) to find a maximum-likelihood
solution to the coadded map at f090 and f150, respectively. Because the
PA4 array had a poor detector yield over the course of the
observation, maps at f220 are treated differently from the other two
frequencies. The resulting excess noise in the ACT f220 maps leads to
a lack of convergence when solving for a coadded map through a maximum
likelihood approach. Therefore, we instead perform a straightforward
inverse-variance weighting in Fourier space to obtain the coadded map
at f220 (see Appendix~\ref{app:f220} for more details).

One caveat in using the Planck HFI maps is that a cosmic infrared background
(CIB) monopole model was deliberately included on a per-frequency
basis due to a lack of sensitivity to the absolute emission level. We
therefore subtracted the CIB monopole in each coadded map following
Table~12 in \cite{Planck_l03:2020}.

\begin{table}[t]
\begin{center}
  \begin{tabular}[t]{cclc}
    \textbf{Band} & \textbf{Planck Dataset} & \textbf{ACT Dataset} & \textbf{Total}\\\hline
    \textbf{f090} & 100~GHz & f090 PA5+PA6 & 6     \\
    \textbf{f150} & 143~GHz & f150 PA4+PA5+PA6 & 8 \\
    \textbf{f220} & 217~GHz & f220 PA4 & 4         \\
  \end{tabular}
  \caption{Subsets of maps coadded at each frequency band. All input maps are two-way
    split maps. The column ``total'' shows the total number of maps
    coadded in each band. For example, 6 different maps went into making
    the f090 coadd map, consisting of two splits from ACT PA4, ACT
    PA5, and Planck 100~GHz, respectively.}
  \label{tab:datasets}
\end{center}
\end{table}

This procedure yields a total of three coadded maps in both temperature
and polarization at f090, f150, and f220, as summarized in
Table~\ref{tab:datasets}. We present a side-by-side comparison between
Planck maps and our three coadded maps in total intensity in
Figure~\ref{fig:t_maps}, and a similar comparison for polarized
intensity for f150 in Figure~\ref{fig:pol_sidebyside}.
It is apparent that the addition of ACT data significantly
improves the angular resolution of the maps in both temperature and
polarization. The coadded polarization maps are presented in
Figure~\ref{fig:QU3bands} in Galactic coordinates. We use the IAU
polarization convention, in which the polarization angle
measures 0$^\circ$ toward Galactic North and increases counter-clockwise \citep{Hamaker:1996}. The ACT Collaboration has adopted the IAU convention for all ACT data products since DR4.
This is in contrast to the COSMO convention \citep{Gorski:2005} adopted in, e.g., the
Planck data releases, that is related to the IAU convention via a sign flip of Stokes $U$, i.e., $U_{\rm COSMO}=-U_{\rm IAU}$. 

A detailed discussion of these maps is presented in Section~\ref{sec:imaps} for total
intensity maps and in Section~\ref{sec:pmaps} for polarization maps.
The final coadded maps have median noise levels of $36\,\mu$K-arcmin at
f090, $33\,\mu$K-arcmin at f150, and $270\,\mu$K-arcmin at f220.

\section{Total intensity maps}
\label{sec:imaps}
Figure~\ref{fig:t_maps} shows the total intensity maps for both Planck-only and the coadded maps for our three frequency bands (f090, f150, f220). 
Many prominent features that were
obscured or unresolved in the Planck maps become apparent with the addition of ACT data, and qualitative changes in map 
morphology with frequency are evident. The Galactic Center
Radio Arc (GCRA), a prominent filament in the Galactic center, is visible at both f090 and
f150 near the center of the coadded maps and to a lesser extent at f220, consistent with it being a 
strong source of synchrotron radiation \cite{Pare:2019}.

The ACT frequency coverage probes a variety of emission mechanisms,
including synchrotron, free-free, thermal dust, and molecular line
emission, at different levels in each of the three bands. To better
visualize the different structures probed at each frequency band, we
combine the coadded maps from three frequency bands into a multicolor
image shown in the upper panel of Figure~\ref{fig:multifreq}. The red,
green, and blue image channels represent the f090, f150, and f220
maps, respectively, after appropriate rescaling. The intensity scaling
(as detailed in the Figure~\ref{fig:multifreq} caption) was chosen
to highlight structures in different bands and to make feature
identification easier. An annotated zoom-in of the three-color intensity map in Figure~\ref{fig:multifreq} is provided in the top panel of Figure~\ref{fig:annot}.

The coherent structures visible in the different colors of
Figures~\ref{fig:multifreq} and \ref{fig:annot} arise from spatial
variations in the relative strengths of the various emission
mechanisms. The radio spectrum of supernova remnants (SNRs) originates
primarily from synchrotron emission \citep{Weiler:1988}, and thus
objects like the SNR candidate G357.7-0.1 (``the Tornado'')
\citep{Milne:1970} and SNR0.9+0.1 \citep{Helfand:1987} appear reddish
yellow in Figure~\ref{fig:multifreq}. Similarly, \added{prominent
  radio sources, including Sgr~A*, the GCRA, and Sgr~B1 \citep[see,
  e.g.,][]{Yusef-Zadeh:1984,Pedlar:1989,Bally:1991}} are strikingly
highlighted in this color, consistent with their strong synchrotron
emission spectrum. Pulsar wind nebulae (PWN), like the Crab Nebula,
also emit highly polarized synchrotron emission with a flat spectral
index \citep{Gaensler:2006}, in contrast to SNRs, which generally emit
synchrotron with a slightly lower polarization fraction and a steeper
spectrum.

Thermal emission from interstellar dust dominates Galactic emission at
far-infrared/submillimeter frequencies. Known molecular \added{cloud complexes like Sgr~B2 \citep[G0.667-0.031; e.g.,][]{Scoville:1975}, Sgr~C \citep[G359.429-0.090; e.g.,][]{Liszt:1995}, and dense} molecular clouds like
the Brick \citep[G0.253+0.016; e.g.,][]{Longmore:2012}, the 20\,km\,s$^{-1}$ Cloud (G359.889-0.093) and 50\,km\,s$^{-1}$ Cloud \citep[G0.070-0.035; e.g.,][]{Gusten:1980}, and 
the Three Little Pigs \citep[G0.145-0.086, G0.106-0.082, and G0.068-0.075; see, e.g.,][for an overview of these molecular clouds]{Battersby:2020} thus appear bright blue/green in Figure~\ref{fig:annot}.

In general, however, the presence of strong molecular line emission in
the CMZ precludes the simple interpretation that low frequencies
correspond to synchrotron emission and high frequencies correspond to
dust emission. Even in the relatively broad Planck and ACT passbands,
line emission can dominate the total intensity in the Galactic center maps.
Indeed, \citet{planck2014-a12} found that 88.6\,GHz HCN emission can alone account for up to 23\,\% of the total intensity in the Planck 
100\,GHz band in this region. CO(1--0) at 115.3\,GHz and CO(2--1) at 230.5\,GHz contribute
significantly to the observed emission in the Planck 100 and 217\,GHz
bands, respectively \citep{planck2014-a12}, while other lines such as HCO$^+$ (89.2\,GHz), CS
(98.0, 147.0, and 244.9\,GHz), $^{13}$CO(1--0) (110.2\,GHz), CN
(113.2, 113.5\,GHz), H$_2$CO (140.8 and 218.2\,GHz), NO
(150.2, 150.5\,GHz), SiO (217.1\,GHz), SO (219.9\,GHz), and
$^{13}$CO(2--1) (220.4\,GHz), among others, are also known to be
present in the Galactic center
\citep[e.g.,][]{Liszt:1978,Sandqvist:1989,Kramer:1998,Lang:2002,Takekawa:2014,Pound:2018,Lu:2021,Schuller:2021}
and will contribute to the observed emission in the ACT and Planck
frequency channels.

The very bright CO(1--0) emission poses a particular challenge for our
analysis, as it falls comfortably within the Planck 100\,GHz passband
but largely outside that of ACT f090
\citep[see][Figure~2]{Naess:2020}. These two frequency channels have
been combined without taking the differences in passbands into
account, leading to CO(1--0) being emphasized on large Planck-dominated
scales in the coadded map, but not on small ACT-dominated scales.
This likely explains the haziness of the emission in purple in
Figure~\ref{fig:multifreq}, where the low-frequency channel (red)
contains significant CO(1--0) emission in the Planck map but is
dominated by other, less prominent emission mechanisms in the ACT map.
A quantitative interpretation of the frequency spectra of particular
regions in the Galactic center will therefore require careful
consideration of bandpass effects, and possibly the use of external
spectroscopic data \citep[e.g.,][]{dame2001, Eden:2020} and/or the CO component maps from Planck \citep{planck2014-a12}. Such spectral
analysis will be the subject of future work, and for now we urge
caution when interpreting the colors in Figure~\ref{fig:multifreq} in
terms of emission mechanisms or spectral indices.

\section{Polarization maps}
\label{sec:pmaps}
\begin{figure*}[ht]
  \centering
  \includegraphics[width=\textwidth]{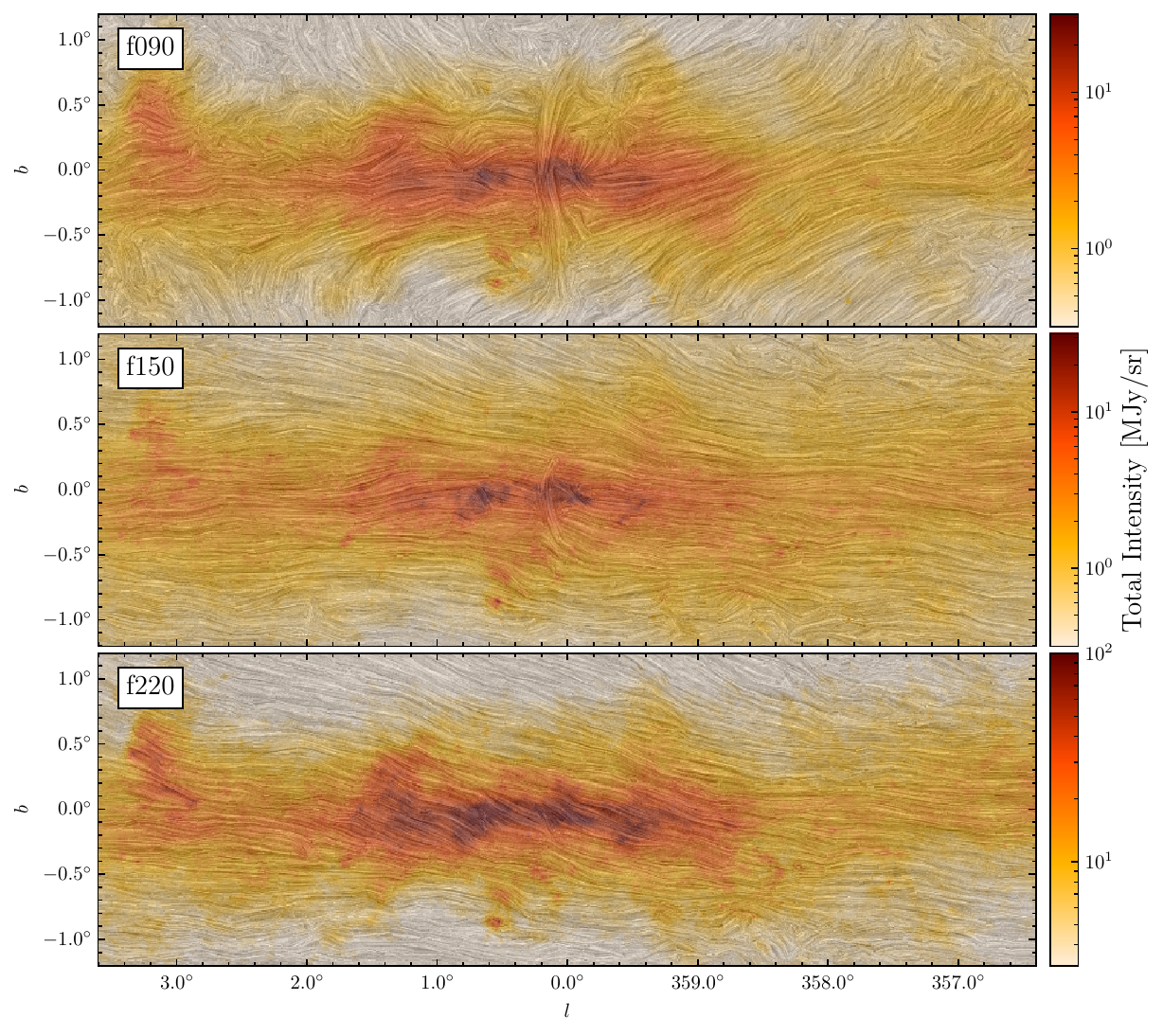}
  \caption{A visualization of magnetic field orientations using
    line-integral-convolution (LIC) with a $1^\circ$ kernel. Contours in
    the map trace magnetic orientations. Rows represent f090, f150,
    and f220 respectively. Total intensity maps are shown in the
    background with the same color scales in
    Figure~\ref{fig:t_maps}.}\label{fig:LIC3band}
\end{figure*}
\begin{figure*}[ht]
  \centering
  \includegraphics[width=0.9\textwidth]{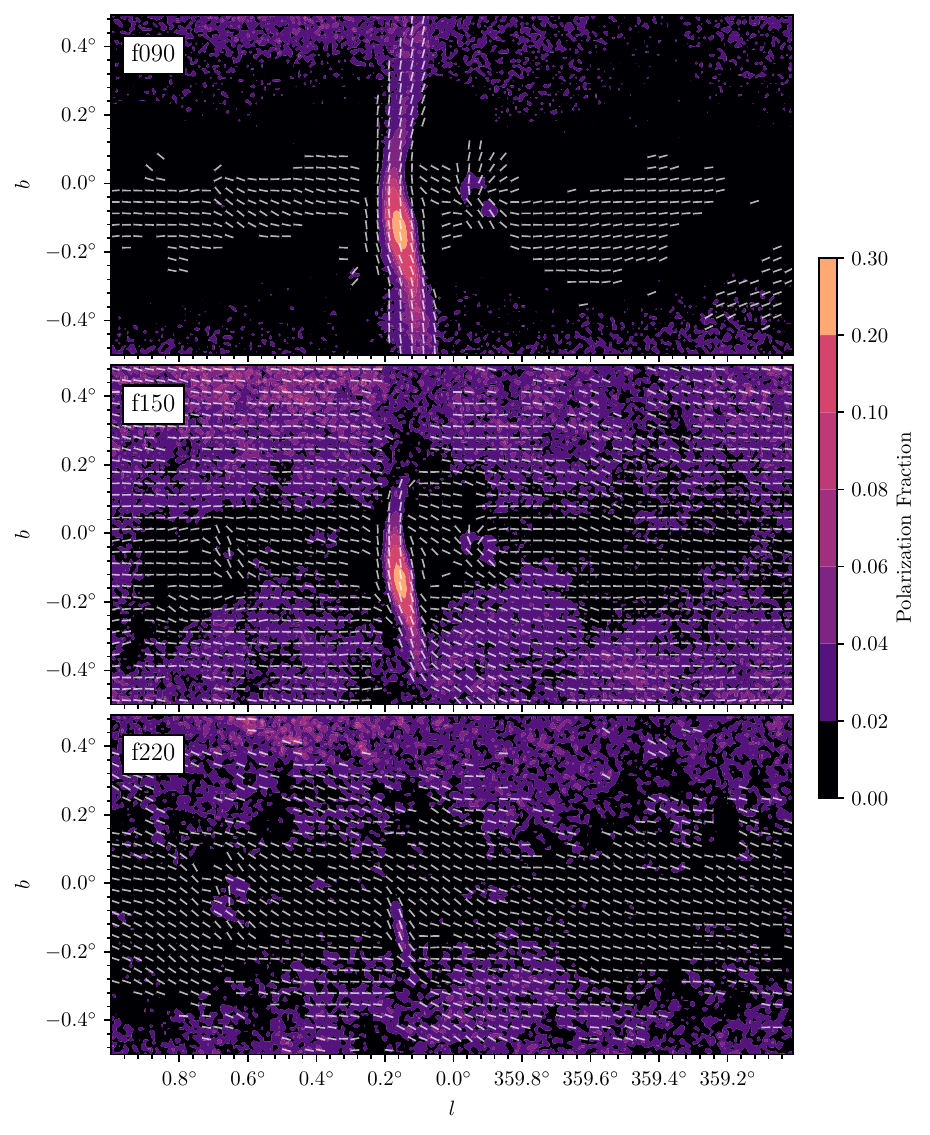}
  \caption{Polarization fractions (background) and magnetic field
    orientation (line segments) are shown for our three bands (f090,
    f150, and f220). To estimate the magnetic field orientations, the
    polarization field is smoothed with a Gaussian kernel FWHM=$2'$,
    and then resampled with a pixel size of $2'$. Line segments with
    large uncertainty in polarization angle $\delta\psi\ge15^\circ$ are masked.}\label{fig:p_psi_3band}
\end{figure*}

Figure~\ref{fig:QU3bands} presents the full-resolution Stokes $Q$ and
$U$ maps obtained through the mapmaking algorithm at each frequency
band. A striking feature of the maps is the strong polarization signal
of the GCRA, extending roughly from $b=-0.5^\circ$ to $b=0.5^\circ$ in both
f090 and f150. The signal is weaker in f220, which is dominated by
polarized dust emission. Strong polarized signals can be generally
seen near the CMZ along the Galactic plane across all frequency bands,
with especially prominent polarization features near regions such as
Sgr~A$^*$ and Sgr~B2.
This suggests that the observed polarization signals are not dominated
by diffuse emission along the LOS, but rather by emission
directly from the CMZ.
Since we are focusing on high S/N regions ($\gtrsim 3$) that are
negligibly impacted by debiasing, we do not debias the polarization
quantities \citep{Plaszczynski:2014}.

To create a three-color polarization image analogous to that in total
intensity, we first compute the polarized intensity $P=\sqrt{Q^2+U^2}$
in each band. We synthesize the three polarized intensity maps into a
three-color image using f090, f150, and f220 as the red, green, and
blue channels, respectively. The result is shown in the lower panel of
Figure~\ref{fig:multifreq}. The polarized emission has a strikingly
different morphology than total intensity (see upper panel of
Figure~\ref{fig:multifreq}). The polarized GCRA stands out
distinctively from the background in red, indicating dominance of
f090, consistent with the prominence of synchrotron radiation in this
region. \added{Similarly, radio sources including Sgr~A$^*$ and
  G359.23–0.82 \citep[``the Mouse''; e.g.,][]{Predehl:1995} appear
  red. On the other hand, molecular cloud complexes such as the
  $l=1.3$ complex \citep{Bally:1988} and G0.55-0.85 \citep[RCW
  142;][]{Gardner:1975} appear blue, consistent with the expected
  predominance of dust emission.}

One quantity of interest is the polarization angle, defined as
\begin{equation}
  \psi = \frac{1}{2} \arctan\left(\frac{U}{Q}\right).
\end{equation}
The polarization angle is directly related to the plane-of-sky
magnetic field orientation by a 90$^\circ$ rotation. Dust grains tend to
align their short axes parallel to the magnetic field, while they
radiate photons preferentially polarized parallel to their long axes.
The synchrotron polarization angle, or electric vector position angle,
is similarly orthogonal to the local magnetic field orientation for
optically thin emission. Hence, the magnetic field orientation is
orthogonal to the polarization angle in both emission mechanisms. We
note, however, that dust and synchrotron emission do not necessarily
trace the same magnetic field, as they generally probe different
volumes along the LOS. The observed magnetic field morphology at a
given frequency depends on the relative contribution of different
emission components, which in turn depends on the spatial distribution
of dust density versus cosmic ray density and the underlying magnetic
field orientation and strength (see \citet{Han:2017} for a review).

Figure~\ref{fig:LIC3band} presents a visualization of the inferred magnetic field orientation in each of our bands using line integral convolution \citep[LIC;][]{Cabral:1993} with
a kernel size of $0.5^\circ$. Each contour in
the map traces the magnetic field orientation. The magnetic
field is approximately parallel to the Galactic plane near the CMZ for
both f090 and f150, and is noticeably tilted for f220 within the range
$\vert l\vert\lesssim 1.5^\circ$. In particular, within a box of
$\vert l\vert < 1.5^\circ$, $\vert b\vert < 0.15^\circ$ we measure the mean polarization angle
to have a tilt of $\simeq 20^\circ$ with respect to the Galactic plane,
consistent with the $\simeq 22^\circ$ tilt previously noted by, e.g.,
PILOT \citep{Mangilli:2019}.

The f090 map is noticeably more disordered, with especially prominent
features at the GCRA, where the plane-of-sky magnetic field is aligned
with the orientation of the arc. This $90^\circ$ flip in polarization
angle at the GCRA has been observed by the QUIET Collaboration
\citep{Ruud:2015} at both 43\,GHz and 97\,GHz. This orthogonal feature
is less prominent at f150 and disappears at f220, as expected from a
synchrotron-dominated signal.

The polarization fraction $p=\sqrt{Q^2+U^2}/I$ in each band is shown in Figure~\ref{fig:p_psi_3band}. In each panel, we overlay the magnetic field orientation in the CMZ at $2'$
resolution. Along the Galactic plane the polarization fraction is generally low, $p \lesssim 2\%$. This is consistent with the previous observations from, e.g., Planck \citep{Planck_l12:2020} and PILOT \citep{Mangilli:2019} that found polarization fractions at the percent level ($\lesssim 1.5\%$) in the Galactic center region. We see coherent magnetic fields even within regions of relatively low polarization fraction, in agreement with both cloud-scale observations and the relatively few wide-area dust polarization measurements, both of which tend to find very ordered magnetic fields \citep{Chuss:2003,Pillai:2015,Mangilli:2019}. The large-scale coherence in the inferred magnetic field direction suggests that the polarized emission is dominated by the CMZ. The low polarization fraction could be due to one of several effects, or to a combination of them. Perhaps the most likely is that the magnetic field orientation fluctuates both along the LOS and within the beam smoothing radius, resulting in depolarization. There are so many emitting regions along the LOS in the Galactic disk that small variations in the magnetic field orientation average out in the LOS integration, such that observed deviations from the mean magnetic field orientation are small. We note, however, that simulations of the Galactic magnetic field used to interpret PILOT data suggest that this effect may not be sufficient on its own to account for the entirety of the observed depolarization \citep{Mangilli:2019}. Another possibility is that the mean field has a significant LOS component. Because magnetically aligned dust grains spin around their short axes, the net dust emission is more strongly polarized for regions with a predominantly plane-of-sky magnetic field than for regions where the magnetic field is more parallel to the LOS.  However, a significant LOS magnetic field component would not be expected to dominate the entirety of the CMZ if the magnetic field has a significant azimuthal component. Finally, it may be that the mean field in the CMZ is itself a product of superimposed, misaligned structures that each have large-scale coherence, e.g., the twisted ring geometry proposed for the distribution of dust density in the CMZ \citep{Molinari:2011}. While possible, such a scenario demands great uniformity in the relative total and polarized intensities in each component to avoid dispersion in the observed polarization angles. On balance, we favor a coherent magnetic field in the CMZ dust, with LOS disorder as the primary driver of low polarization fractions, but more detailed modeling of the present data is warranted to assess the relative importance of each of these effects.

\section{Notable objects}
\label{sec:regions}
\begin{figure*}[ht]
  \centering
  \makebox[\textwidth][c]{\includegraphics[width=1.1\textwidth]{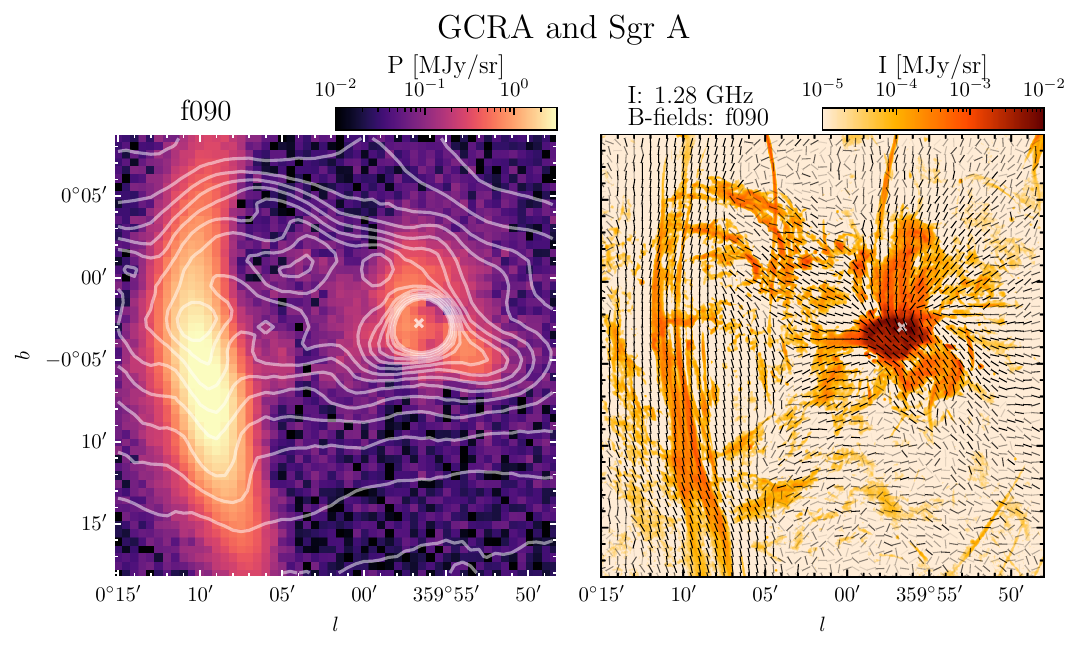}}%
  \caption{GCRA and Sgr~A$^*$. The left panel shows the polarized
    intensity in the region, measured from f090 coadded. Contours show levels of total intensity at f090 with a spacing of 2~MJy\,sr$^{-1}$ up to 30~MJy\,sr$^{-1}$. The right panel shows the inferred magnetic field orientations from the f090 map as line segments in $0.5'$ pixelization (full resolution). Segments are shown with varying opacity that scales linearly with the S/N in polarized intensity and saturates when $S/N=3$. In the
    background we show a radio image of the region from MeerKAT
    \citep{Heywood:2019} which observes at 1.28~GHz in $6''$
    pixelization. The expected location of Sgr~A$^*$ is indicated with a white cross mark in both panels. Note that the MeerKAT image is shown for
    visualization purposes only, as no primary beam corrections have
    been applied, and the entire Galactic plane is seen through
    the primary beam sidelobes. Caution should be taken when
    interpreting the numerical values in this image (see
    \cite{Heywood:2019} for a detailed
    discussion).}\label{fig:SgrAstarB}
\end{figure*}
\begin{figure*}[ht]
  \centering
  \includegraphics[width=1\textwidth]{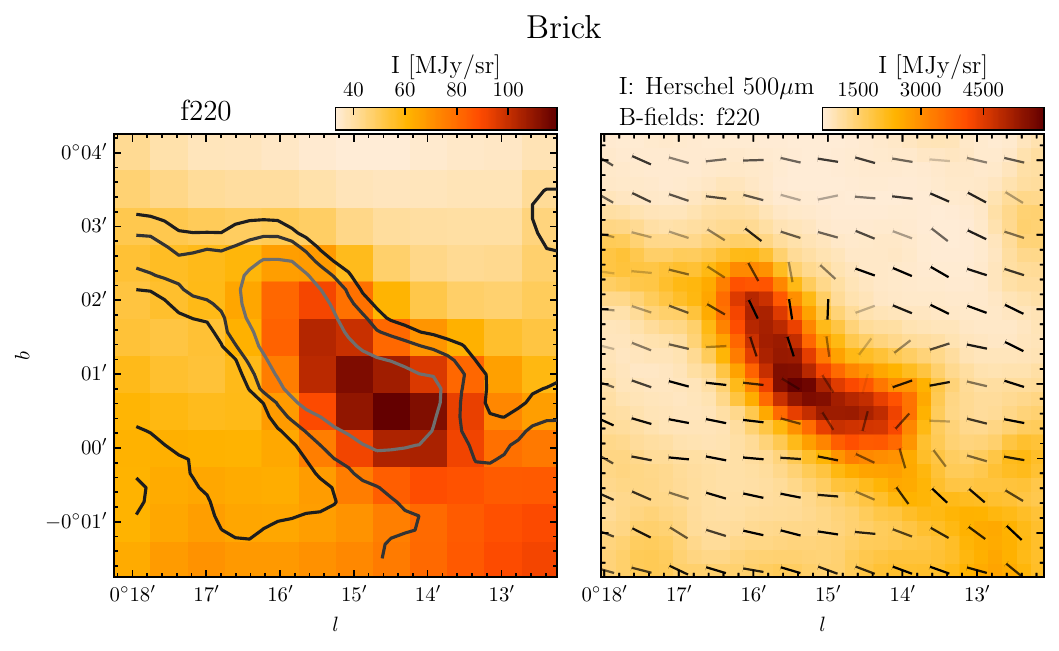}\\%
  \caption{Molecular cloud known as ``the Brick''. \textbf{Left:}
    total intensity measured from ACT+Planck f220 coadd map is plotted
    in the background. The Herschel
    $500\,\mu$m measurements \citep{Molinari:2016} are shown as
    contours indicating 50th, 70th, and 90th percentiles from
    lighter to darker contours. \textbf{Right:} total intensity
    measured by Herschel $500\,\mu$m is shown in the background. We
    show the magnetic field orientation inferred from the f220
    map as line segments. Segments are shown with varying opacity that scales linearly with the S/N in polarized intensity and saturates when $S/N=3$.}\label{fig:brick}
\end{figure*}
\begin{figure*}[ht]
  \centering
  \includegraphics[width=\textwidth]{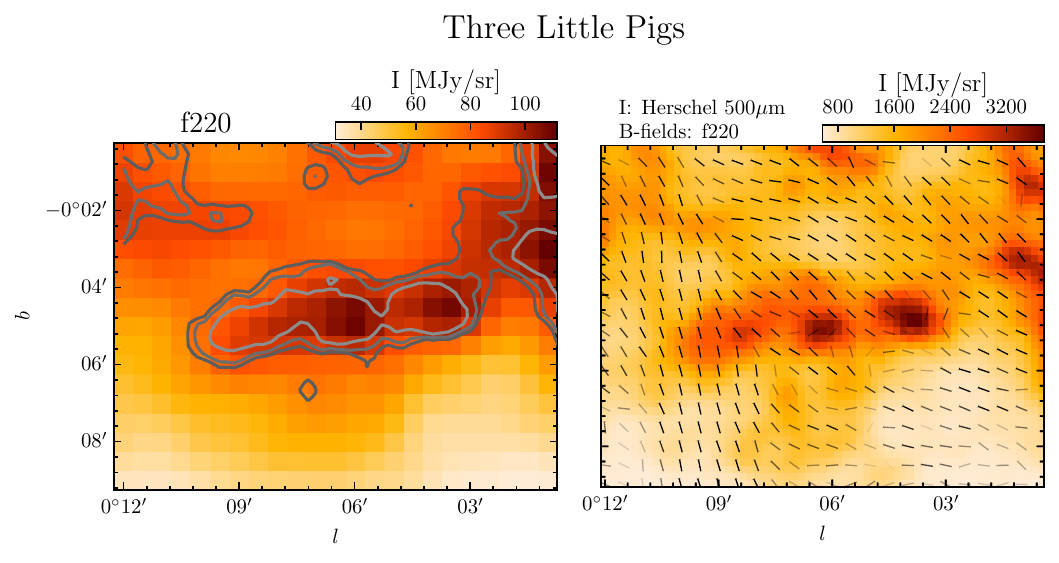}%
  \caption{A cloud triad known as ``the Three Little Pigs'' consisting of G0.145-0.086 (``Straw Cloud''), G0.106-0.082 (``Sticks Cloud''), and G0.068-0.075 (``Stone Cloud''). The data are plotted
    following Figure~\ref{fig:brick}, with the left panel
    showing the ACT+Planck f220 map with the Herschel 500~$\mu$m image
    overlaid as contours (indicating 50th, 70th, and 90th
    percentiles from lighter to darker colors), and the right panel
    showing the Herschel 500~$\mu$m map with the magnetic field
    orientations inferred from the f220 map overlaid as line segments.
    Segments are shown with varying opacity that scales linearly 
    with the S/N in polarized intensity and saturates when S/N$=3$. 
    }\label{fig:3lp}
\end{figure*}
\begin{figure*}[ht]
  \centering
  \includegraphics[width=\textwidth]{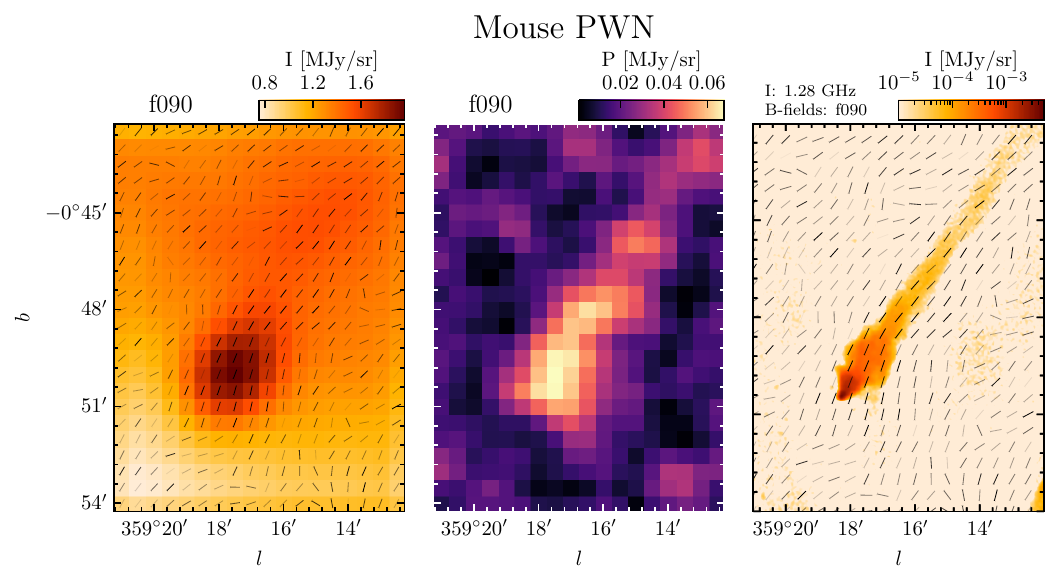}%
  \caption{G359.23–0.82 or ``the Mouse'' is a pulsar wind nebula (PWN)
    traveling with high velocity ($\sim$300 km\,s$^{-1}$) with respect to
    ISM, causing a comet-like tail. The left panel shows the total
    intensity in f090 with magnetic field orientation over-plotted in
    line segments. Both the background and magnetic field are smoothed
    to a resolution of $2.2'$ to increase the signal-to-noise ratio. Segments are shown with varying opacity that scales linearly 
    with the S/N in polarized intensity and saturates when S/N$=3$.
    The middle panel shows the
    polarized intensity in f090 after smoothed to a resolution of $2.2'$. The right panel shows a radio image of the region from MeerKAT
    \citep{Heywood:2019} which observes at 1.28~GHz in $6''$
    pixelization, with the magnetic field orientation from f090 over-plotted
    as line segments similar to the leftmost panel. } \label{fig:mouse}
\end{figure*}
\begin{figure*}[ht]
  \centering
  \includegraphics[width=\textwidth]{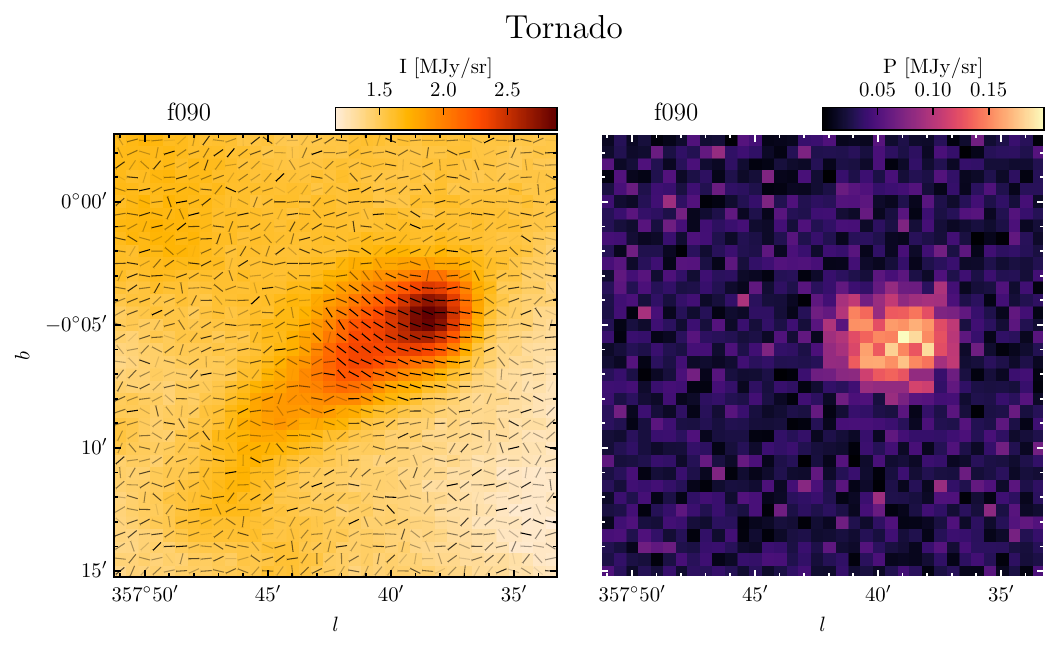}%
  \caption{G357.7-0.1, or ``the Tornado'', is typically classified as
    a supernova remnant. The left plot shows the total intensity in
    its neighborhood in f090 coadded map. Line segments indicate The right plot shows the magnetic field orientations inferred from f090. They are shown with varying opacity that scales linearly with the S/N in polarized intensity and saturates when S/N$=3$. The right panel shows the corresponding polarized intensity map in f090. Both maps are shown at the full resolution from mapmaking ($0.5'$).} \label{fig:tornado}
\end{figure*}
\begin{figure*}[ht]
  \centering
  \includegraphics[width=\textwidth]{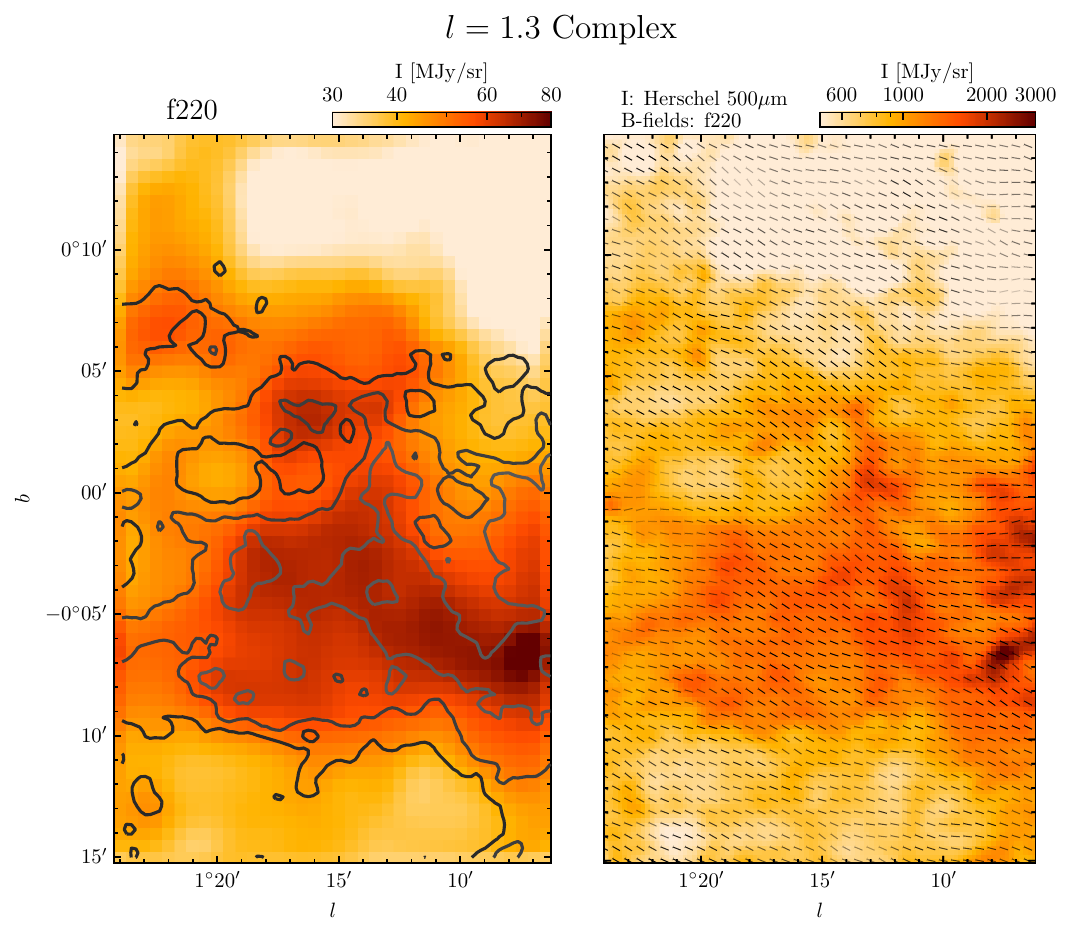}%
  \caption{$l=1.3$ molecular complex. The left plot shows the total
    intensity in f220 (smoothed with FWHM=$1'$) with contours
    indicating the 50th, 70th, 90th percentiles in the Herschel
    500~$\mu$m map. The right plot shows the Herschel 500~$\mu$m map with
    magnetic field orientation inferred from the f220 map as an overlay, after
    smoothed to a resolution of $1.4'$. Segments are shown with varying opacity that scales linearly 
    with the S/N in polarized intensity and saturates when S/N$=3$.} \label{fig:l13}
\end{figure*}

With arcminute resolution in three frequency bands, we detect
many known radio and infrared sources, some of which have not been previously
observed at ACT frequencies. Although the main focus of this paper is
presentation of the Galactic center coadded maps, in this section we demonstrate the
fidelity of these maps and their broad potential for different scientific
investigations by highlighting select objects. All objects discussed in this section are marked in Figure~\ref{fig:annot}, which includes additional selected radio
sources listed in \citet{LaRosa:2000} and submillimeter sources from the CMZoom Survey \citep{Battersby:2020} visible in our maps. This
list of notable sources is non-exhaustive, and in particular, our maps extend
to a wider range in Galactic longitude than either the \citet{LaRosa:2000} or \citet{Battersby:2020}
catalogs.

\subsection{\texorpdfstring{Sgr A and GCRA}{Sgr~A and GCRA}}
Sagittarius A (Sgr~A) is a complex radio source located at the center
of our Galaxy. It consists of Sgr~A East, an extended nonthermal source with a radius of $\sim 3'$, and a thermal source Sgr~A West, which has three-arm spiral morphology and lies within Sgr~A East \citep[e.g.,][]{Ekers:1983,Yusef-Zadeh:1987c,Anantharamaiah:1991}.
Infrared monitoring of stellar orbits in the vicinity of
Sgr~A has also revealed the existence of a supermassive black hole Sgr~A$^*$ 
that lies within Sgr~A West \citep[e.g.,][]{Ghez:2008} and acts as the dynamical center of our Galaxy \citep{Backer:1999}.

The region of sky surrounding Sgr~A$^*$ has been the subject of extensive multifrequency observations both in imaging and polarimetry \citep[e.g.,][]{Stolovy:1996,Bower:1998,Melia:2000,Baganoff:2003,Chuss:2003}.  
Polarized observations in the millimeter bands, in particular, are important for understanding the accretion process near the black hole and associated relativistic emission \citep[e.g.,][]{Agol:2000,Melia:2001}. 
Linear polarization of Sgr~A$^*$ at millimeter wavelengths was first reported by the Submillimetre Common-User Bolometer Array \citep[SCUBA;][]{Aitken:2000}, which they interpret as synchrotron-dominated polarized emission sourced by the gas in the vicinity of the black hole. The observed polarization fraction of Sgr~A$^*$ is $\sim 3\%$ at 2~mm. Subsequent interferometric imaging surveys \citep[e.g.,][]{Macquart:2006, Marrone:2006} measured a $\sim 2\%$ polarization fraction at 3.5~mm, and larger values at higher frequencies.
Strong emission centered on Sgr~A$^*$ is visible in the coadded maps, showing up clearly in the multifrequency image with a yellow color in total intensity (see the upper panel in Figure~\ref{fig:annot}), implying a predominance of synchrotron emission in the region. Its location indicates that the emission is likely dominated by Sgr~A$^*$ itself instead of the overlapping components in Sgr~A that are unresolved with the ACT beam.
Regions surrounding Sgr~A$^*$ are polarized at $2-4\%$ level, as seen in Figure~\ref{fig:p_psi_3band} for f090 and f150, and show up as a reddish ``blob'' in the multifrequency polarimetry (see the lower panel in Figure~\ref{fig:annot}). This may be due to synchrotron emission from the nearby nonthermal filaments within a beam smoothing radius. 
The polarized emission in the vicinity of Sgr~A$^*$ has a lower polarization fraction of $\sim 1.5\%$ at all three bands, consistent with the depolarization noted by SCUBA \citep{Aitken:2000} at 2~mm. The slightly lower polarization fraction seen in the ACT data is likely due to a beam depolarization effect from the larger ACT beam ($\sim 2'$) in comparison to the SCUBA beam ($\sim 34''$ at 2~mm). 

In Figure~\ref{fig:SgrAstarB} we present a zoom-in view of the region surrounding Sgr~A$^*$. The left panel shows the polarized signal in f090 overlaid with contours from the total intensity in f090. Strong emission from Sgr~A$^*$ is seen in total intensity but not in polarization, where the emission is more diffuse and extends $\sim 3'$ away from the central source. This is further evidence that the polarized signal in the vicinity of Sgr~A$^*$ is emitted by the surrounding non-thermal filaments, while the emission from Sgr~A$^*$ itself is highly depolarized. In the right panel we show the inferred magnetic field orientations from the polarized signal at f090 overlaid on top of a radio image of the same region from MeerKAT \citep{Heywood:2019}, which observes at 1.28~GHz with a $6''$ beam. The magnetic field morphology inferred from our f090 map closely follows the underlying non-thermal filamentary structure. The morphology is also in broad agreement with previous Caltech Submillimeter Observatory \citep[CSO;][]{Chuss:2003} observations at a wavelength of 350~$\mu$m with a $20''$ beam.

Figure~\ref{fig:SgrAstarB} also shows the GCRA, a prominent radio feature located at $\sim l=0^\circ 10'$, which consists of a bundle of thin filaments running perpendicular to the Galactic plane \citep[e.g.,][]{Yusef-Zadeh:1987b,Anantharamaiah:1991}. 
The GCRA is known to be a highly polarized synchrotron source, though its origin is still poorly understood. The strong synchrotron emission implies that free electrons are present in the GCRA and are accelerated to relativistic speeds in the presence of a strong magnetic field in the region. Various models have been proposed to explain the source of electrons and the acceleration mechanism \citep[see, e.g.,][for a review]{Serabyn:1994}, though the matter is still under debate.

In millimeter bands, the GCRA has previously been detected at 7~mm \citep{Reich:2000}, 3~mm \citep{Pound:2018}, and 2~mm \citep{Staguhn:2019}, which the latter notes was the highest-frequency detection of the GCRA at the time. Polarized emission from the GCRA has also been previously detected at 2 and 3~mm by \citet{Culverhouse:2011}, and at 3~mm and 7~mm by \citet{Ruud:2015}.
In our coadded maps, GCRA appears in total intensity in both f090 and f150. The associated polarized emission can also be seen clearly in f090 and f150 with polarization fractions reaching $\sim 30\%$. This is considerably higher than the $\sim 10\%$ peak polarization noted by the QUaD Galactic Plane
Survey \citep{Culverhouse:2011} at the same frequencies, likely due to the improved angular resolution in our coadded maps ($2'$ at f090, $1.4'$ at f150) in comparison to \citet{Culverhouse:2011} ($5'$ at 100~GHz, $3.5'$ at 150~GHz). 
The polarized emission from the southern portion of the GCRA is also visible in f220, which is likely the highest frequency at which this structure is detected to date (note especially the f220 $Q$ map in Figure~\ref{fig:QU3bands}). In addition to being fainter at 220 GHz on account of the falling synchrotron spectrum, the GCRA is also obscured by emission from dust along the LOS.
The uniformity of the polarized emission observed in the Arc as seen in Figure~\ref{fig:QU3bands} implies that a highly ordered magnetic field exists along the Arc that deviates sharply from the large-scale magnetic field geometry (see Figure~\ref{fig:LIC3band}). In particular, the magnetic field orientation inferred from f090 (as seen in the right panel of Figure~\ref{fig:SgrAstarB}) aligns closely with the filamentary structure perpendicular to the Galactic plane. This is in broad agreement with the morphology observed at 43~GHz and 96~GHz by QUIET with lower angular resolution \citep{Ruud:2015}.

\subsection{The Brick}
G0.253+0.016, also known as ``the Brick'', is a dense, massive
molecular cloud in the CMZ, and a prominent infrared dark cloud
\citep{Carey:1998, Longmore:2012}. In the context of understanding the
low star formation rate in the Galactic center environment, the Brick
is a particularly interesting case study. Despite its high mass
$(>10^5 M_\sun)$ and density $(> 10^4~\mathrm{cm}^{-3})$, evidence of
star formation is nearly absent in the Brick, and thus it may provide
an ideal opportunity to study the initial conditions of high-mass star
formation \citep{Lis:1994b, Longmore:2012, Kauffmann:2013, Mills:2015,
  Walker:2021}. A number of factors have been invoked to explain the
dearth of star formation in G0.253+0.016, including solenoidal turbulence
driven by strong shear in the CMZ \citep{Federrath:2016,
  Kruijssen:2019, Dale:2019, Henshaw:2019}, or strong cloud-scale
magnetic fields \citep[$B \sim \mathrm{mG}$,][]{Pillai:2015}.

The Brick stands out at high contrast to the background in the coadded
total intensity maps at both 150 and 220\,GHz. Our polarization
measurements at these frequencies probe the magnetic field structure
in the dust toward G0.253+0.016 at $\sim$arcminute scales. These
observations complement $20''$ resolution polarization data at 350
$\mu$m from the Caltech Submilimeter Observatory
\citep[CSO;][]{Dotson:2010, Pillai:2015}. We find that the inferred
magnetic field orientation is aligned parallel to the long axis of the
Brick on the plane of the sky (Figure~\ref{fig:brick}), and the
polarization angles are very ordered in this region, in agreement with
the CSO data at smaller angular scales. \citet{Pillai:2015} use the
strong coherence of the magnetic field orientation in the Brick to
compare the inferred magnetic field strength to the gas velocity
dispersion measured from $\mathrm{N_2 H ^+}$ emission
\citep{Kauffmann:2013}. Those authors find that magnetic fields
dominate over turbulence in the Brick. The coherent magnetic field
structure in our observations is consistent with the expectation that
turbulence in the Brick is sub-Alfv\'enic at the scales probed by ACT.
The ACT polarized emission is brightest at the northern part of the
Brick, with a peak polarization fraction of $1.8\%$. The polarized
intensity is lower in the southern portion of the cloud, and the SNR
on the polarized intensity drops below 3. This depolarization may be
due in part to unresolved polarization structure within the ACT beam,
and/or to incoherent contributions to the polarized emission along the
LOS.

\subsection{The Three Little Pigs}
The cloud triad G0.145-0.086, G0.106-0.082, and G0.068-0.075 visible
in Figure~\ref{fig:3lp} has been dubbed ``the Three Little Pigs.'' All three clouds have been noted as
a set of compact dusty sources in the CMZoom Survey \citep{Battersby:2020},
while G0.068-0.075 also appears in the SCUBA-2 Compact Source Catalog
\citep{Parsons:2018}. As Figure~\ref{fig:3lp} illustrates, each cloud is also apparent in the 500\,$\mu$m data from Herschel Infrared Galactic Plane Survey Herschel \citep[Hi-GAL;][]{Molinari:2016}. Interestingly, the $3''$ resolution 230\,GHz
observations with the Submillimeter Array as part of the CMZoom Survey
have revealed a dearth of substructure in G0.145-0.086 (``Straw
Cloud''), somewhat more substructure in G0.106-0.082 (``Sticks
Cloud''), and yet more in G0.068-0.075 (``Stone Cloud'').

ACT f220 measurements give a first look at the magnetic field geometry
in these clouds at arcminute resolution. The Straw Cloud, perhaps
owing to a lower column density or lack of dense substructure, has a magnetic field orientation that deviates little from 
the large-scale field structure. In contrast,
both the Sticks and Stone Clouds have polarization angles in their
interiors that are highly misaligned with the large-scale magnetic field. Similar to the depolarization observed toward the Brick, the
cancellation of polarized emission from dust in different regions
within the cloud and/or other dust along the LOS may explain
the low polarized intensities observed, particularly in the Stone
Cloud.

\subsection{The Mouse}
G359.23–0.82, also known as ``the Mouse'', is a PWN powered by the
young X-ray source PSR~J1747–2958 \citep{Predehl:1995, Camilo:2002}.
G359.23–0.82 was originally discovered in radio continuum data from
the Very Large Array (VLA), and derives its nickname from its bright
compact nebula ``head'' and extended radio ``tail''
\citep{Yusef-Zadeh:1987a}. The Mouse is strongly linearly polarized at
centimeter wavelengths \citep{Yusef-Zadeh:2005}. Distances to
PSR~J1747–2958 and the Mouse are uncertain, but they are not at the
Galactic center: observations of neutral hydrogen absorption set the
maximum distance to G359.23–0.82 at $\sim5.5$ kpc \citep{Uchida:1992}.
\citet{Gaensler:2004} argue for a distance of $\sim5$\,kpc, a value now
commonly adopted \citep[e.g.,][]{Klingler:2018}. At 5\,kpc, the
transverse velocity of PSR~J1747–2958 is $306 \pm 43$\,km\,s$^{-1}$
\citep{Hales:2009}. The Mouse is a striking example of a bow shock
nebula, formed by the interaction of the pulsar with the ambient ISM
as it travels at supersonic speeds \citep[e.g.,][]{Gaensler:2006}.

The Mouse is a prominent object in the ACT f090 map, both in total and
polarized intensity (Figure~\ref{fig:mouse}). In particular, polarized emission is detected significantly across the peak of the Mouse, which is expected for a PWN. Significant polarized emission is also detected along its tail, and exhibits a similar morphology as seen by MeerKAT at 1.28~GHz \citep{Heywood:2019} with a $6''$ beam, albeit at lower resolution in the ACT data.
The implied magnetic field orientation in the f090 band is roughly parallel to the Mouse's extended tail, consistent with observations at 3.5 and 6~cm by the VLA \citep{Yusef-Zadeh:2005}. 
The Mouse is traveling eastward in declination, which
is roughly toward the lower left-hand corner of Figure~\ref{fig:mouse}.

\subsection{The Tornado}
G357.7-0.1, ``the Tornado,'' is typically classified as a SNR, though
its unusual properties have prevented a definitive explanation
\citep{Gaensler:2003,Chawner:2020}. The Tornado has been long observed
in radio imaging and polarimetry
\citep[e.g.,][]{Milne:1970,Shaver:1985,Law:2008}, which consistently
show a bright ``head'' region and a ``tail'' region roughly $10'$ in
extent. Recently, mid- and far-infrared dust emission has been
detected with Spitzer and Herschel, revealing a large dust reservoir
in the head region ($\sim 17$\,M$_\odot$) and consistent with interstellar
matter swept up in a supernova blast wave \citep{Chawner:2020}. The
head of the Tornado has also been detected by Chandra in X-rays
without evidence for embedded point sources \citep{Gaensler:2003},
lending further credence to its classification as a SNR.
However, the provenance of the tail is still unresolved
\citep[see][for a recent discussion]{Chawner:2020}.

The Tornado is prominent in the f090 and f150 Stokes $Q$ and $U$ maps,
but not f220 (see Figure~\ref{fig:QU3bands}). Likewise, the region
stands out in reddish brown in the three-color polarization map
(Figure~\ref{fig:annot}). This suggests the prominence of
synchrotron emission in this source. A closer examination of the
Tornado in the f090 band is presented in Figure~\ref{fig:tornado}.
Here, we see the extended tail region in total intensity but not in
polarization, while the head is prominent in both. 
This morphology is consistent with 4.9\,GHz polarimetric observations by \citet{Shaver:1985}. The inferred magnetic field at f090 is approximately perpendicular to the extended tail in the eastern side of the Tornado, and is tilted toward the head on the western side. This is also in a broad agreement with the magnetic field morphology noted by \citep{Shaver:1985} at 4.9~GHz.  
We observe a maximum polarization fraction of the Tornado in f090 
of $8.5\%\pm 1\%$,
slightly lower than the $\sim 10\%$ observed at 4.9\,GHz at significantly higher resolution \citep[$12\times26''$ beam,][]{Shaver:1985}. It is likely that much of the difference is due to more beam depolarization in the ACT data.

\subsection{\texorpdfstring{$l=1.3$ Complex}{l=1.3 Complex}}
The combination of ACT and Planck data used in the coadded maps enables large regions to be mapped with fidelity on both large and small angular scales. Likewise, the high sensitivity of the polarimetry permits mapping of more diffuse regions of molecular clouds, not just bright cores. These capabilities are highlighted in the $20\times30'$ maps of the $l = 1.3$ complex in Figure~\ref{fig:l13}.

The $l=1.3$ complex is a large, high-velocity-dispersion molecular cloud complex extending from roughly 1.2--1.6$^\circ$ in Galactic longitude \citep{Bally:1988}. The elevated abundance of SiO and high ratio of CO(3--2) to CO(1--0) emission in some clouds within the complex suggest the presence of strong shocks, perhaps from cloud-cloud collisions or supernova explosions \citep{Huettemeister:1998,Oka:2001,Rodriguez-Fernandez:2006,Tanaka:2007,Parsons:2018,Tsujimoto:2021}. This complex may sit at the intersection of a dust lane with the nuclear ring, supplying it with material \citep{Huettemeister:1998,Fux:1999,Rodriguez-Fernandez:2006,Liszt:2008}.

Total emission from the $l = 1.3$ in f220 and Herschel 500\,$\mu$m \citep{Molinari:2016} is presented in Figure~\ref{fig:l13}, with good morphological correspondence between the two maps. In the right panel, we overlay the f220 magnetic field orientation on the higher resolution Herschel map. While many density structures show clear alignment with the magnetic field orientation, this is not universally observed. The highest intensity regions have comparatively low polarized intensities, suggesting elevated magnetic field disorder or a loss of grain alignment in the densest regions.

\section{Summary and future prospects}\label{sec:conclusion}
We have presented new arcminute-resolution maps of the Galactic
center region at microwave frequencies by combining data from ACT and
Planck. Known radio features appear at high significance in both total
intensity and polarization in three frequency bands. The polarization
maps provide a frequency-dependent probe of magnetic fields,
demonstrating a change in the observed magnetic field morphology as the fractional
contributions of synchrotron radiation and thermal dust emission from different regions within the Galactic center along the LOS vary
with frequency. With wide-field maps at higher angular resolution, 
we identified
known radio sources and molecular clouds, some of which have not
previously been observed in polarization at microwave frequencies.
With three frequency bands, our total intensity maps reveal the rich
physical environment in the CMZ with spatially varying combinations of
different emission mechanisms, including synchrotron, free-free, dust,
and molecular line emission in the CMZ. Separation of these emission components
will be the subject of a follow-up paper.

The coadded maps produced in this work \replaced{will be made}{are now} publicly available on \replaced{LAMBDA}{the NASA Legacy Archive Microwave Background Data Analysis \citep[LAMBDA;][]{Miller:2018}\footnote{\url{https://lambda.gsfc.nasa.gov/product/act/actadv_sr_gc_1_info.cfm}}}. These maps are suitable for tracing magnetic field morphology across the Galactic center region and measuring the total and polarized emission from individual sources. However, caution is urged for multifrequency analyses due to the bandpass mismatch between ACT and Planck that results in a slight scale-dependence of effective band centers for different emission mechanisms. As discussed in Section~\ref{sec:imaps}, CO(1-0) emission falls within the Planck 100 GHz passband but not f090, amplifying bandpass mismatch effects in the resulting coadded map.

ACT has continued to observe the Galactic center during 2020,
collecting a similar amount of data to that used in this work. In
addition, the daytime data from both 2019 and 2020 can, in principle,
be corrected for thermal telescope distortions \citep{Aiola:2020},
which would again double the total amount of data. Therefore, ACT maps
with half the pixel noise variance of those presented here are possible based
solely on data that has already been collected. Additionally, we plan to apply
the mapping techniques used here to approximately 70 degrees of the
Galactic plane covered by ACT from 2019 and 2020. Furthermore, the 
addition of the low-frequency array to ACT in 2020 \citep{Li:2018,Simon:2018} will also 
allow us to map the Galactic plane at 27\,GHz and 39\,GHz, likely
yielding new insights on the Galactic center environment. 

The next observational step at these frequencies will be the Large
Aperture Telescope of the Simons Observatory \citep{Ade:2019},
anticipated to see first light in 2023 from the same site in Chile.
This new instrument will have the same 6-meter diameter primary as
ACT, but with an instrumented focal plane of 5 times larger area \citep{Zhu:2021}. The
nominal scan strategy will continuously cover the entire sky in the
declination range between $+25^\circ$ and $-40^\circ$, providing coverage of
over 100 degrees of the Galactic plane in five frequency bands in both
total intensity and polarization. The five-year map noise should
improve on ACT by roughly a factor of three. 
The Galactic center will be observed at higher frequencies by the CCAT-prime project \citep{Choi:2020} and will also be a good target for future balloon-borne instruments, which can achieve sub-arcminute resolution with similar sensitivity at even higher frequencies, e.g., BLAST Observatory \citep{Lowe:2020}.
By 2030, we can also anticipate data from CMB-S4 \citep{Abazajian:2016}, with an additional map noise improvement by a factor of four. This unrivaled combination of resolution, sky coverage, and sensitivity at microwave frequencies will enable many new inquiries into the properties of the Milky Way. 

\begin{acknowledgments}
This work was supported by the U.S. National Science Foundation through awards
AST-0408698, AST-0965625, and AST-1440226 for the ACT project, as well as
awards PHY-0355328, PHY-0855887 and PHY-1214379. Funding was also provided by
Princeton University, the University of Pennsylvania, and a Canada Foundation
for Innovation (CFI) award to UBC. ACT operates in the Parque Astron\'omico
Atacama in northern Chile under the auspices of the Comisi\'on Nacional de
Investigaci\'on (CONICYT). 

Computations were performed using Princeton Research Computing
resources at Princeton University, the Niagara supercomputer at the
SciNet HPC Consortium and on the Simons-Popeye cluster of the Flatiron
Institute. SciNet is funded by the CFI under the auspices of Compute
Canada, the Government of Ontario, the Ontario Research Fund---Research
Excellence, and the University of Toronto. The development of
multichroic detectors and lenses was supported by NASA grants
NNX13AE56G and NNX14AB58G. Detector research at NIST was supported by
the NIST Innovations in Measurement Science program. Research at
Perimeter Institute is supported in part by the Government of Canada
through the Department of Innovation, Science and Industry Canada and
by the Province of Ontario through the Ministry of Colleges and
Universities.

S.E.C. acknowledges support by the Friends of the Institute for
Advanced Study Membership. EC acknowledges support from the STFC
Ernest Rutherford Fellowship ST/M004856/2 and STFC Consolidated Grant
ST/S00033X/1, and from the European Research Council (ERC) under the
European Union’s Horizon 2020 research and innovation programme (Grant
agreement No. 849169). SKC acknowledges support from NSF award
AST-2001866. R.D. thanks CONICYT for grant BASAL CATA AFB-170002.
J.P.H. would like to thank Fernando Camilo and Ian Heywood for
providing the MeerKAT image from their 2019 Nature paper in digital
form; he also acknowledges financial support from NASA grant
NNX15AK71G and NSF Astronomy and Astrophysics Research Grant number
AST-1615657 to Rutgers University. E.S. is supported by the
Chamberlain fellowship at Lawrence Berkeley National Laboratory. NS
acknowledge support from NSF grant number AST-1907657. Zhilei Xu is
supported by the Gordon and Betty Moore Foundation. We thank Bruce
Draine for spotting typos in the manuscript.

\end{acknowledgments}
\vspace{4mm}
\appendix

\section{Calibration method}
\label{app:calibration}
In ACT DR4 \citep{Aiola:2020}, the data were calibrated from raw data
acquisition units to physical units with
\begin{equation}
  d^{\rm pW}=d^{\rm DAQ}\times R_{\rm BS}\times g_{\rm atm}\times f, 
\end{equation}
where $d^{\rm pW}$, $d^{\rm DAQ}$ represent detector data in physical
unit and data acquisition unit respectively, $R_{\rm BS}$ represents
the intrinsic responsivity of each detector measured from the most
recent bias step, $g_{\rm atm}$ is the atmospheric correction factor
and $f$ represents an optical flatfield. Both $g_{\rm atm}$ and $f$
are estimated from detector responsiveness to the atmospheric signal
which is treated as a common mode for all detectors. In the presence
of non-atmospheric thermal contamination signatures this approach to
estimating the common mode may bias the calibration. Preliminary
analyses on the Advanced ACT data collected after 2017 have shown evidences
for the presence of such thermal contamination. Hence, the calibration
method needs to be updated to account for this bias.

To circumvent the problem, we switch from a common-mode based
calibration to a planet based approach, given by
\begin{equation}
  d^{\rm pW}=d^{\rm DAQ}\times R_{\rm BS}\times f_{\rm p},
\end{equation}
where we have dropped the atmospheric correction $g_{\rm atm}$.
$f_{\rm p}$ now represents an optical flatfield measured from detector
responsiveness to emission from Uranus instead of the atmosphere.
This leads to an improved calibration model and better
gain stability, and this is expected to be the standard calibration
method for future data releases of ACT.

One caveat to note is that ACT DR4 (and earlier) maps are 
calibrated to Planck maps in a final step. This step, however,
is missing here due to the preliminary nature of the Advanced ACT 
data in 2019, but since the Planck calibration is found to be 
close to the planet calibrations for ACT DR4, and we are using an improved 
calibration model from DR4, we do not expect this to be a concern 
and estimate a $O(1\%)$ uncertainty in global gain calibration as a result.

\section{Beam leakage correction}\label{app:beam}
\begin{figure*}[t]
  \centering
  \makebox[\textwidth][c]{\includegraphics[width=1.15\textwidth]{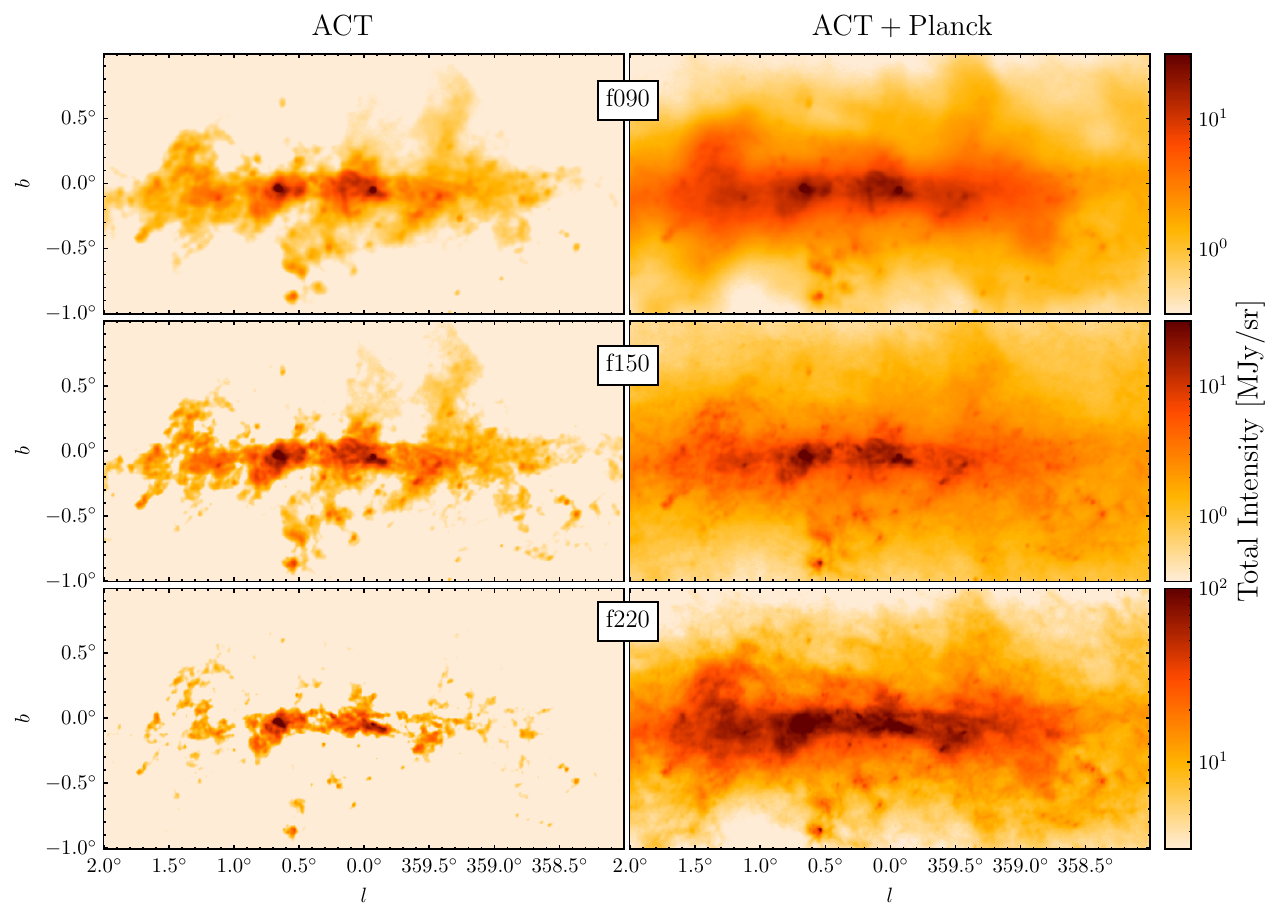}}
  \caption{\added{Comparison between ACT-only maps (left column)
      and ACT+Planck coadded maps (right column) in total intensity,
      similar to Figure~\ref{fig:t_maps}. Rows from top to bottom
      correspond to f090, f150, and f220, respectively. The ACT maps
      shown here are produced by coadding maps from different detector
      arrays with inverse variance weightings at each frequency band
      respectively. Each map is plotted on a logarithmic color scale
      from 0.3--30 MJy\,sr$^{-1}$ for f090 and f150, and from from
      3--100 MJy\,sr$^{-1}$ for f220.}} \label{fig:t_maps_act}
\end{figure*}

To reduce the contamination in the polarization maps caused by beam
leakage effects, we modeled the observed polarization maps
$P_{\rm obs}$ as a sum of a beam-convolved sky map $P_{\rm sky}$ and a
leakage component, given by
\begin{equation}
  P_{\rm obs}(\mathbf{x}) = P_{\rm sky}(\mathbf{x}) + \int \frac{d^2k}{(2\pi)^2}~\tilde{T}_{\rm sky}(\mathbf{k}) \tilde{B}_{\rm eff}(\mathbf{k}) e^{-i\mathbf{k}\cdot\mathbf{x}},
\end{equation}
where the leakage component is a convolution of the beam-convolved
temperature map $T_{\rm sky}$ with an effective beam $B_{\rm eff}$
given in Fourier space by
\begin{equation}
  \tilde{B}_{\rm eff}(\mathbf{k})\equiv \tilde{B}_T^{-1}(\mathbf{k})\tilde{B}_P(\mathbf{k})
  ~~~.
\end{equation}
This represents a combination of deconvolving the temperature beam
$B_{T}$ and convolving with a leakage beam $B_{P}$. Our strategy is
then to build a model of $\tilde{B}_{\rm eff}(\mathbf{k})$, convolve
it with the temperature map, and deproject it from the observed
polarization maps for each dataset.

As we assumed Uranus signal is unpolarized, any signal we measure in
the polarization is a sign of T-to-P leakage. Hence, we modeled
$\tilde{B}_{\rm eff}$ using observations of Uranus made in the same
observation season, with the following steps:
\begin{enumerate}
\item We made Uranus planet maps for each Uranus observation in a
  source-centered reference frame with the scan direction as the
  horizontal axis.
\item We reprojected each planet map into the Galactic coordinate
  system. As the $Q$/$U$ reference frame needs to be rotated depending on
  the scan directions, and the Galactic center observations consist of two scan
  directions taken during rising and setting respectively, we
  accounted for the difference in scan directions by rotating the $Q$/$U$
  reference frame for both rising and setting scan directions and
  performed a weighted average (for each planet map) depending on the
  total number of rising and setting scans made during Galactic center  
  observations.
\item We stacked all reprojected Uranus maps for each dataset (per
  frequency band per array) with inverse-variance weighting to obtain
  estimates of $B_T$ and $B_P$ for each dataset.
\item We performed a real space cut to remove noise outside a radius
  of $r_{\rm max}$ in $B_T$ and $B_P$ for each dataset, then
  calculated $\tilde{B}_{\rm eff}(\mathbf{k})=\tilde{B}_P/\tilde{B}_T$
  in Fourier space.
\item We further cleaned $\tilde{B}_{\rm eff}(\mathbf{k})$ with a
  $k$-space cut $k\le k_{\rm max}$ to get rid of small scale noise in
  the beam model. We then re-filled the $k$-space outside $k_{\rm max}$
  by mirroring the value at $k=k_{\rm max}$ with
  $\tilde{B}_{\rm eff}(\mathbf{k})\vert_{k>k_{\rm max}} =
  \tilde{B}_{\rm eff}(\mathbf{k}~k_{\rm max}/k)$. In practice, the
  details of the extrapolating function make little difference. This
  specific function is chosen to ensure that the transition at
  $k=k_{\rm max}$ is smooth, and it extends naturally to infinity.
\end{enumerate}
In Table~\ref{tab:beamcuts}, we listed the choices of $r_{\rm max}$
and $k_{\rm max}$ for each dataset. As a result of these steps, we
obtained $\tilde{B}_{\rm eff}(\mathbf{k})$ for each dataset
respectively. Treating the coadded temperature map as the ``true''
sky model $T_{\rm sky}$ at each frequency band, we predicted the
expected T-to-P leakage from the derived leakage beam model
$\tilde{B}_{\rm eff}(\mathbf{k})$ and subsequently deprojected the
expected leakage from each observed map. \added{The resulting ACT 
maps after leakage correction are shown in Figure~\ref{fig:t_maps_act} 
in comparison to the coadded maps.}
\begin{table}
  \centering
  \begin{tabular}[t]{cccc}
    array & freq & $r_{\rm max}$ & $k_{\rm max}$\\\hline
    PA4 & f150 & 5$'$ & 14500\\
    PA4 & f220 & 4$'$ & 20000\\
    PA5 & f090 & 8$'$ & 10500\\
    PA5 & f150 & 5$'$ & 16500\\
    PA6 & f090 & 8$'$ & 10500\\
    PA6 & f150 & 5$'$ & 16500\\
  \end{tabular}
  \caption{Parameters used when building a 2D leakage beam model (in
    step 4 and step 5).}
  \label{tab:beamcuts}
\end{table}

\section{Coadded f220 maps}
\label{app:f220}
Maps at f220 are much noisier than the two other bands, so using the
same coadding pipeline results in a slow convergence. Instead, we
adopted a simpler Fourier-based coadding algorithm. Denote the ACT
f220 map as $m_1$, and Planck 217\,GHz map as $m_2$. The coadd map
$m_{\rm coadd}$ was obtained with a simple inverse-variance coadding
as
\begin{equation}
  m_{\rm coadd} = \left(N_1^{-1}+N_2^{-1}\right)^{-1}\left(N_1^{-1}m_1 + N_2^{-1}m_2\right),
\end{equation}
where $N_1$ and $N_2$ are noise covariance matrices assumed for ACT
and Planck maps respectively. Specifically, we assumed simple
Fourier-based noise model with
$N_1=w_1^2[1+(\ell/\ell_{\rm knee})^{-\alpha}]$ and $N_2=w_2^2 b_1/b_2$, with
$w_1=268.97~\mu$K$'$, $w_2=124.22~\mu$K$'$ the noise levels of ACT and
Planck maps respectively, $\ell_{\rm knee}=4000$, and $\alpha=-3$. $b_1$ and
$b_2$ represent the beam model in ACT and Planck respectively, and the
factor $b_1/b_2$ in $N_2$ represents the effect of a combination of
deconvolution of the Planck beam and convolution with the ACT beam in the
Planck noise model. In addition, we further applied a high-pass
Butterworth filter in the ACT map with $\ell_{c}=2500$ and $\alpha=-5$ along
the two cross-linked scan directions with a width given by the beam
FWHM in Fourier space. This helps suppress excess noise
along the scan directions. We then applied an additional high-pass
Butterworth filter with $\ell_{c}=200$ and $\alpha=-10$ to suppress the
large-scale atmospheric noise in the ACT f220 map. Both of these
filters are included in the ACT noise model $N_\ell$.

The inverse-variance map of the output coadd map was estimated using the
same weighted average from Planck inverse-variance map ${\cal N}_1^{-1}$ and 
ACT inverse-variance map ${\cal N}_2^{-1}$, given by
\begin{equation}
  {\cal N}_{\rm coadd} = \overline{c}_1^2{\cal N}_1+\overline{c}_2^2{\cal N}_2
\end{equation}
with $\overline{c}_1$, $\overline{c}_2$ defined as the $\ell$-space mean of
$c_{1,2}\equiv N_{1,2}^{-1}/(N_1^{-1}+N_2^{-1})$ from $\ell=2500$ to
$5000$ where we expect to be dominated by white noise.

\bibliography{references,GClibrary,Planck_bib}{}

\begin{thebibliography}{}
\expandafter\ifx\csname natexlab\endcsname\relax\def\natexlab#1{#1}\fi
\providecommand{\url}[1]{\href{#1}{#1}}
\providecommand{\dodoi}[1]{doi:~\href{http://doi.org/#1}{\nolinkurl{#1}}}
\providecommand{\doeprint}[1]{\href{http://ascl.net/#1}{\nolinkurl{http://ascl.net/#1}}}
\providecommand{\doarXiv}[1]{\href{https://arxiv.org/abs/#1}{\nolinkurl{https://arxiv.org/abs/#1}}}

\bibitem[{{Ade} {et~al.}(2019){Ade}, {Aguirre}, {Ahmed}, {Aiola}, {Ali},
  {Alonso}, {Alvarez}, {Arnold}, {Ashton}, {Austermann}, \& et~al.}]{Ade:2019}
{Ade}, P., {Aguirre}, J., {Ahmed}, Z., {et~al.} 2019, \jcap, 2019, 056,
  \dodoi{10.1088/1475-7516/2019/02/056}

\bibitem[{{Agol}(2000)}]{Agol:2000}
{Agol}, E. 2000, \apjl, 538, L121, \dodoi{10.1086/312818}

\bibitem[{{Aiola} {et~al.}(2020){Aiola}, {Calabrese}, {Maurin}, {Naess},
  {Schmitt}, {Abitbol}, {Addison}, {Ade}, {Alonso}, {Amiri}, {Amodeo},
  {Angile}, {Austermann}, {Baildon}, {Battaglia}, {Beall}, {Bean}, {Becker},
  {Bond}, {Bruno}, {Calafut}, {Campusano}, {Carrero}, {Chesmore}, {Cho},
  {Choi}, {Clark}, {Cothard}, {Crichton}, {Crowley}, {Darwish}, {Datta},
  {Denison}, {Devlin}, {Duell}, {Duff}, {Duivenvoorden}, {Dunkley},
  {D{\"u}nner}, {Essinger-Hileman}, {Fankhanel}, {Ferraro}, {Fox}, {Fuzia},
  {Gallardo}, {Gluscevic}, {Golec}, {Grace}, {Gralla}, {Guan}, {Hall},
  {Halpern}, {Han}, {Hargrave}, {Hasselfield}, {Helton}, {Henderson},
  {Hensley}, {Hill}, {Hilton}, {Hilton}, {Hincks}, {Hlo{\v{z}}ek}, {Ho},
  {Hubmayr}, {Huffenberger}, {Hughes}, {Infante}, {Irwin}, {Jackson}, {Klein},
  {Knowles}, {Koopman}, {Kosowsky}, {Lakey}, {Li}, {Li}, {Li}, {Lokken},
  {Louis}, {Lungu}, {MacInnis}, {Madhavacheril}, {Maldonado}, {Mallaby-Kay},
  {Marsden}, {McMahon}, {Menanteau}, {Moodley}, {Morton}, {Namikawa}, {Nati},
  {Newburgh}, {Nibarger}, {Nicola}, {Niemack}, {Nolta}, {Orlowski-Sherer},
  {Page}, {Pappas}, {Partridge}, {Phakathi}, {Pisano}, {Prince}, {Puddu}, {Qu},
  {Rivera}, {Robertson}, {Rojas}, {Salatino}, {Schaan}, {Schillaci}, {Sehgal},
  {Sherwin}, {Sierra}, {Sievers}, {Sifon}, {Sikhosana}, {Simon}, {Spergel},
  {Staggs}, {Stevens}, {Storer}, {Sunder}, {Switzer}, {Thorne}, {Thornton},
  {Trac}, {Treu}, {Tucker}, {Vale}, {Van Engelen}, {Van Lanen}, {Vavagiakis},
  {Wagoner}, {Wang}, {Ward}, {Wollack}, {Xu}, {Zago}, \& {Zhu}}]{Aiola:2020}
{Aiola}, S., {Calabrese}, E., {Maurin}, L., {et~al.} 2020, \jcap, 2020, 047,
  \dodoi{10.1088/1475-7516/2020/12/047}

\bibitem[{{Aitken} {et~al.}(2000){Aitken}, {Greaves}, {Chrysostomou},
  {Jenness}, {Holland}, {Hough}, {Pierce-Price}, \& {Richer}}]{Aitken:2000}
{Aitken}, D.~K., {Greaves}, J., {Chrysostomou}, A., {et~al.} 2000, \apjl, 534,
  L173, \dodoi{10.1086/312685}

\bibitem[{{Anantharamaiah} {et~al.}(1991){Anantharamaiah}, {Pedlar}, {Ekers},
  \& {Goss}}]{Anantharamaiah:1991}
{Anantharamaiah}, K.~R., {Pedlar}, A., {Ekers}, R.~D., \& {Goss}, W.~M. 1991,
  \mnras, 249, 262, \dodoi{10.1093/mnras/249.2.262}

\bibitem[{{Backer} \& {Sramek}(1999)}]{Backer:1999}
{Backer}, D.~C., \& {Sramek}, R.~A. 1999, \apj, 524, 805,
  \dodoi{10.1086/307857}

\bibitem[{{Baganoff} {et~al.}(2003){Baganoff}, {Maeda}, {Morris}, {Bautz},
  {Brandt}, {Cui}, {Doty}, {Feigelson}, {Garmire}, {Pravdo}, {Ricker}, \&
  {Townsley}}]{Baganoff:2003}
{Baganoff}, F.~K., {Maeda}, Y., {Morris}, M., {et~al.} 2003, \apj, 591, 891,
  \dodoi{10.1086/375145}

\bibitem[{{Bally} {et~al.}(1991){Bally}, {Casement}, {Goss}, {Lindqvist},
  {Nyman}, {Mehringer}, {Palmer}, \& {Yusef-Zadeh}}]{Bally:1991}
{Bally}, J., {Casement}, S., {Goss}, W.~M., {et~al.} 1991, in Bulletin of the
  American Astronomical Society, Vol.~23, 890

\bibitem[{{Bally} {et~al.}(1988){Bally}, {Stark}, {Wilson}, \&
  {Henkel}}]{Bally:1988}
{Bally}, J., {Stark}, A.~A., {Wilson}, R.~W., \& {Henkel}, C. 1988, \apj, 324,
  223, \dodoi{10.1086/165891}

\bibitem[{{Barnes} {et~al.}(2017){Barnes}, {Longmore}, {Battersby}, {Bally},
  {Kruijssen}, {Henshaw}, \& {Walker}}]{Barnes:2017}
{Barnes}, A.~T., {Longmore}, S.~N., {Battersby}, C., {et~al.} 2017, \mnras,
  469, 2263, \dodoi{10.1093/mnras/stx941}

\bibitem[{{Battersby} {et~al.}(2020){Battersby}, {Keto}, {Walker}, {Barnes},
  {Callanan}, {Ginsburg}, {Hatchfield}, {Henshaw}, {Kauffmann}, {Kruijssen},
  {Longmore}, {Lu}, {Mills}, {Pillai}, {Zhang}, {Bally}, {Butterfield},
  {Contreras}, {Ho}, {Ott}, {Patel}, \& {Tolls}}]{Battersby:2020}
{Battersby}, C., {Keto}, E., {Walker}, D., {et~al.} 2020, \apjs, 249, 35,
  \dodoi{10.3847/1538-4365/aba18e}

\bibitem[{{Bennett} {et~al.}(2013){Bennett}, {Larson}, {Weiland}, {Jarosik},
  {Hinshaw}, {Odegard}, {Smith}, {Hill}, {Gold}, {Halpern}, {Komatsu}, {Nolta},
  {Page}, {Spergel}, {Wollack}, {Dunkley}, {Kogut}, {Limon}, {Meyer}, {Tucker},
  \& {Wright}}]{Bennett:2013}
{Bennett}, C.~L., {Larson}, D., {Weiland}, J.~L., {et~al.} 2013, \apjs, 208,
  20, \dodoi{10.1088/0067-0049/208/2/20}

\bibitem[{{Bower} \& {Backer}(1998)}]{Bower:1998}
{Bower}, G.~C., \& {Backer}, D.~C. 1998, \apjl, 496, L97,
  \dodoi{10.1086/311259}

\bibitem[{Cabral \& Leedom(1993)}]{Cabral:1993}
Cabral, B., \& Leedom, L.~C. 1993, in Proceedings of the 20th annual conference
  on Computer graphics and interactive techniques, 263--270

\bibitem[{{Camilo} {et~al.}(2002){Camilo}, {Manchester}, {Gaensler}, \&
  {Lorimer}}]{Camilo:2002}
{Camilo}, F., {Manchester}, R.~N., {Gaensler}, B.~M., \& {Lorimer}, D.~R. 2002,
  \apjl, 579, L25, \dodoi{10.1086/344832}

\bibitem[{{Carey} {et~al.}(1998){Carey}, {Clark}, {Egan}, {Price}, {Shipman},
  \& {Kuchar}}]{Carey:1998}
{Carey}, S.~J., {Clark}, F.~O., {Egan}, M.~P., {et~al.} 1998, \apj, 508, 721,
  \dodoi{10.1086/306438}

\bibitem[{{Chawner} {et~al.}(2020){Chawner}, {Howard}, {Gomez}, {Matsuura},
  {Priestley}, {Barlow}, {De Looze}, {Papageorgiou}, {Marsh}, {Smith},
  {Noriega-Crespo}, {Rho}, \& {Dunne}}]{Chawner:2020}
{Chawner}, H., {Howard}, A.~D.~P., {Gomez}, H.~L., {et~al.} 2020, \mnras, 499,
  5665, \dodoi{10.1093/mnras/staa2925}

\bibitem[{{Choi} {et~al.}(2018){Choi}, {Austermann}, {Beall}, {Crowley},
  {Datta}, {Duff}, {Gallardo}, {Ho}, {Hubmayr}, {Koopman}, {Li}, {Nati},
  {Niemack}, {Page}, {Salatino}, {Simon}, {Staggs}, {Stevens}, {Ullom}, \&
  {Wollack}}]{choi/etal:2018}
{Choi}, S.~K., {Austermann}, J., {Beall}, J.~A., {et~al.} 2018, Journal of Low
  Temperature Physics, 193, 267, \dodoi{10.1007/s10909-018-1982-4}

\bibitem[{{Choi} {et~al.}(2020){Choi}, {Austermann}, {Basu}, {Battaglia},
  {Bertoldi}, {Chung}, {Cothard}, {Duff}, {Duell}, {Gallardo}, {Gao}, {Herter},
  {Hubmayr}, {Niemack}, {Nikola}, {Riechers}, {Rossi}, {Stacey}, {Stevens},
  {Vavagiakis}, {Vissers}, \& {Walker}}]{Choi:2020}
{Choi}, S.~K., {Austermann}, J., {Basu}, K., {et~al.} 2020, Journal of Low
  Temperature Physics, 199, 1089, \dodoi{10.1007/s10909-020-02428-z}

\bibitem[{{Chuss} {et~al.}(2003){Chuss}, {Davidson}, {Dotson}, {Dowell},
  {Hildebrand}, {Novak}, \& {Vaillancourt}}]{Chuss:2003}
{Chuss}, D.~T., {Davidson}, J.~A., {Dotson}, J.~L., {et~al.} 2003, \apj, 599,
  1116, \dodoi{10.1086/379538}

\bibitem[{{CMB-S4 Collaboration} {et~al.}(2016){CMB-S4 Collaboration},
  {Abazajian}, {Adshead}, {Ahmed}, {Allen}, {Alonso}, {Arnold}, {Baccigalupi},
  {Bartlett}, {Battaglia}, {Benson}, {Bischoff}, {Borrill}, {Buza},
  {Calabrese}, {Caldwell}, {Carlstrom}, {Chang}, {Crawford}, {Cyr-Racine}, {De
  Bernardis}, {de Haan}, {di Serego Alighieri}, {Dunkley}, {Dvorkin}, {Errard},
  {Fabbian}, {Feeney}, {Ferraro}, {Filippini}, {Flauger}, {Fuller},
  {Gluscevic}, {Green}, {Grin}, {Grohs}, {Henning}, {Hill}, {Hlozek}, {Holder},
  {Holzapfel}, {Hu}, {Huffenberger}, {Keskitalo}, {Knox}, {Kosowsky}, {Kovac},
  {Kovetz}, {Kuo}, {Kusaka}, {Le Jeune}, {Lee}, {Lilley}, {Loverde},
  {Madhavacheril}, {Mantz}, {Marsh}, {McMahon}, {Meerburg}, {Meyers}, {Miller},
  {Munoz}, {Nguyen}, {Niemack}, {Peloso}, {Peloton}, {Pogosian}, {Pryke},
  {Raveri}, {Reichardt}, {Rocha}, {Rotti}, {Schaan}, {Schmittfull}, {Scott},
  {Sehgal}, {Shandera}, {Sherwin}, {Smith}, {Sorbo}, {Starkman}, {Story}, {van
  Engelen}, {Vieira}, {Watson}, {Whitehorn}, \& {Kimmy Wu}}]{Abazajian:2016}
{CMB-S4 Collaboration}, {Abazajian}, K.~N., {Adshead}, P., {et~al.} 2016, arXiv
  e-prints, arXiv:1610.02743.
\newblock \doarXiv{1610.02743}

\bibitem[{{Crutcher} {et~al.}(1996){Crutcher}, {Roberts}, {Mehringer}, \&
  {Troland}}]{Crutcher:1996}
{Crutcher}, R.~M., {Roberts}, D.~A., {Mehringer}, D.~M., \& {Troland}, T.~H.
  1996, \apjl, 462, L79, \dodoi{10.1086/310031}

\bibitem[{{Culverhouse} {et~al.}(2011){Culverhouse}, {Ade}, {Bock}, {Bowden},
  {Brown}, {Cahill}, {Castro}, {Church}, {Friedman}, {Ganga}, {Gear}, {Gupta},
  {Hinderks}, {Kovac}, {Lange}, {Leitch}, {Melhuish}, {Memari}, {Murphy},
  {Orlando}, {Pryke}, {Schwarz}, {O'Sullivan}, {Piccirillo}, {Rajguru},
  {Rusholme}, {Taylor}, {Thompson}, {Turner}, {Wu}, {Zemcov}, \& {QUaD
  Collaboration}}]{Culverhouse:2011}
{Culverhouse}, T., {Ade}, P., {Bock}, J., {et~al.} 2011, \apjs, 195, 8,
  \dodoi{10.1088/0067-0049/195/1/8}

\bibitem[{{Dale} {et~al.}(2019){Dale}, {Kruijssen}, \& {Longmore}}]{Dale:2019}
{Dale}, J.~E., {Kruijssen}, J.~M.~D., \& {Longmore}, S.~N. 2019, \mnras, 486,
  3307, \dodoi{10.1093/mnras/stz888}

\bibitem[{{Dame} {et~al.}(2001){Dame}, {Hartmann}, \& {Thaddeus}}]{dame2001}
{Dame}, T.~M., {Hartmann}, D., \& {Thaddeus}, P. 2001, \apj, 547, 792,
  \dodoi{10.1086/318388}

\bibitem[{{Dotson} {et~al.}(2010){Dotson}, {Vaillancourt}, {Kirby}, {Dowell},
  {Hildebrand}, \& {Davidson}}]{Dotson:2010}
{Dotson}, J.~L., {Vaillancourt}, J.~E., {Kirby}, L., {et~al.} 2010, \apjs, 186,
  406, \dodoi{10.1088/0067-0049/186/2/406}

\bibitem[{{Eden} {et~al.}(2020){Eden}, {Moore}, {Currie}, {Rigby},
  {Rosolowsky}, {Su}, {Kim}, {Parsons}, {Morata}, {Chen}, {Minamidani}, {Park},
  {Ragan}, {Urquhart}, {Rani}, {Tahani}, {Billington}, {Deb}, {Figura},
  {Fujiyoshi}, {Joncas}, {Liao}, {Liu}, {Ma}, {Tuan-Anh}, {Yun}, {Zhang},
  {Zhu}, {Henshaw}, {Longmore}, {Kobayashi}, {Thompson}, {Ao},
  {Campbell-White}, {Ching}, {Chung}, {Duarte-Cabral}, {Fich}, {Gao}, {Graves},
  {Jiang}, {Kemper}, {Kuan}, {Kwon}, {Lee}, {Lee}, {Liu}, {Pe{\~n}aloza},
  {Peretto}, {Phuong}, {Pineda}, {Plume}, {Puspitaningrum}, {Samal}, {Soam},
  {Sun}, {Tang}, {Traficante}, {White}, {Yan}, {Yang}, {Yuan}, {Yue}, {Bemis},
  {Brunt}, {Chen}, {Cho}, {Clark}, {Cyganowski}, {Friberg}, {Fuller}, {Han},
  {Hoare}, {Izumi}, {Kim}, {Kim}, {Kim}, {Koch}, {Kuno}, {Lacialle}, {Lai},
  {Lee}, {Lee}, {Li}, {Liu}, {Mairs}, {Pan}, {Qian}, {Scicluna}, {Shi}, {Shi},
  {Srinivasan}, {Tan}, {Thomas}, {Torii}, {Trejo}, {Umemoto}, {Violino},
  {Wallstr{\"o}m}, {Wang}, {Wu}, {Yuan}, {Zhang}, {Zhang}, {Zhou}, \&
  {Zhou}}]{Eden:2020}
{Eden}, D.~J., {Moore}, T.~J.~T., {Currie}, M.~J., {et~al.} 2020, \mnras, 498,
  5936, \dodoi{10.1093/mnras/staa2734}

\bibitem[{{Ekers} {et~al.}(1983){Ekers}, {van Gorkom}, {Schwarz}, \&
  {Goss}}]{Ekers:1983}
{Ekers}, R.~D., {van Gorkom}, J.~H., {Schwarz}, U.~J., \& {Goss}, W.~M. 1983,
  \aap, 122, 143

\bibitem[{{Federrath} {et~al.}(2016){Federrath}, {Rathborne}, {Longmore},
  {Kruijssen}, {Bally}, {Contreras}, {Crocker}, {Garay}, {Jackson}, {Testi}, \&
  {Walsh}}]{Federrath:2016}
{Federrath}, C., {Rathborne}, J.~M., {Longmore}, S.~N., {et~al.} 2016, \apj,
  832, 143, \dodoi{10.3847/0004-637X/832/2/143}

\bibitem[{{Federrath} {et~al.}(2017){Federrath}, {Rathborne}, {Longmore},
  {Kruijssen}, {Bally}, {Contreras}, {Crocker}, {Garay}, {Jackson}, {Testi}, \&
  {Walsh}}]{Federrath:2017}
{Federrath}, C., {Rathborne}, J.~M., {Longmore}, S.~N., {et~al.} 2017, in The
  Multi-Messenger Astrophysics of the Galactic Centre, ed. R.~M. {Crocker},
  S.~N. {Longmore}, \& G.~V. {Bicknell}, Vol. 322, 123--128,
  \dodoi{10.1017/S1743921316012357}

\bibitem[{{Ferri{\`e}re}(2009)}]{Ferriere:2009}
{Ferri{\`e}re}, K. 2009, \aap, 505, 1183, \dodoi{10.1051/0004-6361/200912617}

\bibitem[{{Ferri{\`e}re}(2011)}]{Ferriere:2011}
{Ferri{\`e}re}, K. 2011, in Astrophysical Dynamics: From Stars to Galaxies, ed.
  N.~H. {Brummell}, A.~S. {Brun}, M.~S. {Miesch}, \& Y.~{Ponty}, Vol. 271,
  170--178, \dodoi{10.1017/S1743921311017583}

\bibitem[{{Ferri{\`e}re} {et~al.}(2007){Ferri{\`e}re}, {Gillard}, \&
  {Jean}}]{Ferriere:2007}
{Ferri{\`e}re}, K., {Gillard}, W., \& {Jean}, P. 2007, \aap, 467, 611,
  \dodoi{10.1051/0004-6361:20066992}

\bibitem[{{Fowler} {et~al.}(2007){Fowler}, {Niemack}, {Dicker}, {Aboobaker},
  {Ade}, {Battistelli}, {Devlin}, {Fisher}, {Halpern}, {Hargrave}, {Hincks},
  {Kaul}, {Klein}, {Lau}, {Limon}, {Marriage}, {Mauskopf}, {Page}, {Staggs},
  {Swetz}, {Switzer}, {Thornton}, \& {Tucker}}]{Fowler:2007}
{Fowler}, J.~W., {Niemack}, M.~D., {Dicker}, S.~R., {et~al.} 2007, \ao, 46,
  3444, \dodoi{10.1364/AO.46.003444}

\bibitem[{{Fux}(1999)}]{Fux:1999}
{Fux}, R. 1999, \aap, 345, 787.
\newblock \doarXiv{astro-ph/9903154}

\bibitem[{{Gaensler} {et~al.}(2003){Gaensler}, {Fogel}, {Slane}, {Miller},
  {Wijnands}, {Eikenberry}, \& {Lewin}}]{Gaensler:2003}
{Gaensler}, B.~M., {Fogel}, J.~K.~J., {Slane}, P.~O., {et~al.} 2003, \apjl,
  594, L35, \dodoi{10.1086/378260}

\bibitem[{{Gaensler} \& {Slane}(2006)}]{Gaensler:2006}
{Gaensler}, B.~M., \& {Slane}, P.~O. 2006, \araa, 44, 17,
  \dodoi{10.1146/annurev.astro.44.051905.092528}

\bibitem[{{Gaensler} {et~al.}(2004){Gaensler}, {van der Swaluw}, {Camilo},
  {Kaspi}, {Baganoff}, {Yusef-Zadeh}, \& {Manchester}}]{Gaensler:2004}
{Gaensler}, B.~M., {van der Swaluw}, E., {Camilo}, F., {et~al.} 2004, \apj,
  616, 383, \dodoi{10.1086/424906}

\bibitem[{{Gardner} \& {Whiteoak}(1975)}]{Gardner:1975}
{Gardner}, F.~F., \& {Whiteoak}, J.~B. 1975, \mnras, 171, 29P,
  \dodoi{10.1093/mnras/171.1.29P}

\bibitem[{{Ghez} {et~al.}(2008){Ghez}, {Salim}, {Weinberg}, {Lu}, {Do}, {Dunn},
  {Matthews}, {Morris}, {Yelda}, {Becklin}, {Kremenek}, {Milosavljevic}, \&
  {Naiman}}]{Ghez:2008}
{Ghez}, A.~M., {Salim}, S., {Weinberg}, N.~N., {et~al.} 2008, \apj, 689, 1044,
  \dodoi{10.1086/592738}

\bibitem[{{Ginsburg} {et~al.}(2019){Ginsburg}, {Mills}, {Battersby},
  {Longmore}, \& {Kruijssen}}]{Ginsburg:2019}
{Ginsburg}, A., {Mills}, E. A.~C., {Battersby}, C.~D., {Longmore}, S.~N., \&
  {Kruijssen}, J.~M.~D. 2019, \baas, 51, 220.
\newblock \doarXiv{1903.04525}

\bibitem[{{G{\'o}rski} {et~al.}(2005){G{\'o}rski}, {Hivon}, {Banday},
  {Wandelt}, {Hansen}, {Reinecke}, \& {Bartelmann}}]{Gorski:2005}
{G{\'o}rski}, K.~M., {Hivon}, E., {Banday}, A.~J., {et~al.} 2005, \apj, 622,
  759, \dodoi{10.1086/427976}

\bibitem[{{G{\"u}sten}(1989)}]{Gusten:1989}
{G{\"u}sten}, R. 1989, in The Center of the Galaxy, ed. M.~{Morris}, Vol. 136,
  89

\bibitem[{{G{\"u}sten} \& {Downes}(1980)}]{Gusten:1980}
{G{\"u}sten}, R., \& {Downes}, D. 1980, \aap, 87, 6

\bibitem[{{Hales} {et~al.}(2009){Hales}, {Gaensler}, {Chatterjee}, {van der
  Swaluw}, \& {Camilo}}]{Hales:2009}
{Hales}, C.~A., {Gaensler}, B.~M., {Chatterjee}, S., {van der Swaluw}, E., \&
  {Camilo}, F. 2009, \apj, 706, 1316, \dodoi{10.1088/0004-637X/706/2/1316}

\bibitem[{{Hamaker} \& {Bregman}(1996)}]{Hamaker:1996}
{Hamaker}, J.~P., \& {Bregman}, J.~D. 1996, \aaps, 117, 161

\bibitem[{{Han}(2017)}]{Han:2017}
{Han}, J.~L. 2017, \araa, 55, 111, \dodoi{10.1146/annurev-astro-091916-055221}

\bibitem[{{Helfand} \& {Becker}(1987)}]{Helfand:1987}
{Helfand}, D.~J., \& {Becker}, R.~H. 1987, \apj, 314, 203,
  \dodoi{10.1086/165050}

\bibitem[{{Henderson} {et~al.}(2016){Henderson}, {Allison}, {Austermann},
  {Baildon}, {Battaglia}, {Beall}, {Becker}, {De Bernardis}, {Bond},
  {Calabrese}, {Choi}, {Coughlin}, {Crowley}, {Datta}, {Devlin}, {Duff},
  {Dunkley}, {D{\"u}nner}, {van Engelen}, {Gallardo}, {Grace}, {Hasselfield},
  {Hills}, {Hilton}, {Hincks}, {Hloẑek}, {Ho}, {Hubmayr}, {Huffenberger},
  {Hughes}, {Irwin}, {Koopman}, {Kosowsky}, {Li}, {McMahon}, {Munson}, {Nati},
  {Newburgh}, {Niemack}, {Niraula}, {Page}, {Pappas}, {Salatino}, {Schillaci},
  {Schmitt}, {Sehgal}, {Sherwin}, {Sievers}, {Simon}, {Spergel}, {Staggs},
  {Stevens}, {Thornton}, {Van Lanen}, {Vavagiakis}, {Ward}, \&
  {Wollack}}]{henderson/etal:2016}
{Henderson}, S.~W., {Allison}, R., {Austermann}, J., {et~al.} 2016, Journal of
  Low Temperature Physics, 184, 772, \dodoi{10.1007/s10909-016-1575-z}

\bibitem[{{Henshaw} {et~al.}(2019){Henshaw}, {Ginsburg}, {Haworth}, {Longmore},
  {Kruijssen}, {Mills}, {Sokolov}, {Walker}, {Barnes}, {Contreras}, {Bally},
  {Battersby}, {Beuther}, {Butterfield}, {Dale}, {Henning}, {Jackson},
  {Kauffmann}, {Pillai}, {Ragan}, {Riener}, \& {Zhang}}]{Henshaw:2019}
{Henshaw}, J.~D., {Ginsburg}, A., {Haworth}, T.~J., {et~al.} 2019, \mnras, 485,
  2457, \dodoi{10.1093/mnras/stz471}

\bibitem[{{Heywood} {et~al.}(2019){Heywood}, {Camilo}, {Cotton}, {Yusef-Zadeh},
  {Abbott}, {Adam}, {Aldera}, {Bauermeister}, {Booth}, {Botha}, {Botha},
  {Brederode}, {Brits}, {Buchner}, {Burger}, {Chalmers}, {Cheetham}, {de
  Villiers}, {Dikgale-Mahlakoana}, {du Toit}, {Esterhuyse}, {Fanaroff},
  {Foley}, {Fourie}, {Gamatham}, {Goedhart}, {Gounden}, {Hlakola}, {Hoek},
  {Hokwana}, {Horn}, {Horrell}, {Hugo}, {Isaacson}, {Jonas}, {Jordaan},
  {Joubert}, {J{\'o}zsa}, {Julie}, {Kapp}, {Kenyon}, {Kotz{\'e}}, {Kriel},
  {Kusel}, {Lehmensiek}, {Liebenberg}, {Loots}, {Lord}, {Lunsky}, {Macfarlane},
  {Magnus}, {Magozore}, {Mahgoub}, {Main}, {Malan}, {Malgas}, {Manley},
  {Maree}, {Merry}, {Millenaar}, {Mnyandu}, {Moeng}, {Monama}, {Mphego}, {New},
  {Ngcebetsha}, {Oozeer}, {Otto}, {Passmoor}, {Patel}, {Peens-Hough},
  {Perkins}, {Ratcliffe}, {Renil}, {Rust}, {Salie}, {Schwardt}, {Serylak},
  {Siebrits}, {Sirothia}, {Smirnov}, {Sofeya}, {Swart}, {Tasse}, {Taylor},
  {Theron}, {Thorat}, {Tiplady}, {Tshongweni}, {van Balla}, {van der Byl}, {van
  der Merwe}, {van Dyk}, {Van Rooyen}, {Van Tonder}, {Van Wyk}, {Wallace},
  {Welz}, \& {Williams}}]{Heywood:2019}
{Heywood}, I., {Camilo}, F., {Cotton}, W.~D., {et~al.} 2019, \nat, 573, 235,
  \dodoi{10.1038/s41586-019-1532-5}

\bibitem[{{Ho} {et~al.}(2017){Ho}, {Austermann}, {Beall}, {Choi}, {Cothard},
  {Crowley}, , {Datta}, {Duff}, {Gallardo}, {Hasselfield}, {Henderson},
  {Koopman}, {Niemack}, {Salatino}, {Simon}, {Staggs}, \& {Wollack}}]{ho/2017}
{Ho}, S.-P.~P., {Austermann}, J.~A., {Beall}, J.~A., {et~al.} 2017, in
  \procspie, Vol. 9914, Millimeter, Submillimeter, and Far-Infrared Detectors
  and Instrumentation for Astronomy VIII, 991418, \dodoi{10.1117/12.2233113}

\bibitem[{{Huettemeister} {et~al.}(1998){Huettemeister}, {Dahmen},
  {Mauersberger}, {Henkel}, {Wilson}, \& {Martin-Pintado}}]{Huettemeister:1998}
{Huettemeister}, S., {Dahmen}, G., {Mauersberger}, R., {et~al.} 1998, \aap,
  334, 646.
\newblock \doarXiv{astro-ph/9803054}

\bibitem[{{Kauffmann} {et~al.}(2013){Kauffmann}, {Pillai}, \&
  {Zhang}}]{Kauffmann:2013}
{Kauffmann}, J., {Pillai}, T., \& {Zhang}, Q. 2013, \apjl, 765, L35,
  \dodoi{10.1088/2041-8205/765/2/L35}

\bibitem[{{Klingler} {et~al.}(2018){Klingler}, {Kargaltsev}, {Pavlov}, {Ng},
  {Beniamini}, \& {Volkov}}]{Klingler:2018}
{Klingler}, N., {Kargaltsev}, O., {Pavlov}, G.~G., {et~al.} 2018, \apj, 861, 5,
  \dodoi{10.3847/1538-4357/aac6e0}

\bibitem[{{Kramer} {et~al.}(1998){Kramer}, {Staguhn}, {Ungerechts}, \&
  {Sievers}}]{Kramer:1998}
{Kramer}, C., {Staguhn}, J., {Ungerechts}, H., \& {Sievers}, A. 1998, in The
  Central Regions of the Galaxy and Galaxies, ed. Y.~{Sofue}, Vol. 184, 173

\bibitem[{{Kruijssen} \& {Longmore}(2013)}]{Kruijssen:2013}
{Kruijssen}, J.~M.~D., \& {Longmore}, S.~N. 2013, \mnras, 435, 2598,
  \dodoi{10.1093/mnras/stt1634}

\bibitem[{{Kruijssen} {et~al.}(2014){Kruijssen}, {Longmore}, {Elmegreen},
  {Murray}, {Bally}, {Testi}, \& {Kennicutt}}]{Kruijssen:2014}
{Kruijssen}, J.~M.~D., {Longmore}, S.~N., {Elmegreen}, B.~G., {et~al.} 2014,
  \mnras, 440, 3370, \dodoi{10.1093/mnras/stu494}

\bibitem[{{Kruijssen} {et~al.}(2019){Kruijssen}, {Dale}, {Longmore}, {Walker},
  {Henshaw}, {Jeffreson}, {Petkova}, {Ginsburg}, {Barnes}, {Battersby},
  {Immer}, {Jackson}, {Keto}, {Krieger}, {Mills}, {S{\'a}nchez-Monge},
  {Schmiedeke}, {Suri}, \& {Zhang}}]{Kruijssen:2019}
{Kruijssen}, J.~M.~D., {Dale}, J.~E., {Longmore}, S.~N., {et~al.} 2019, \mnras,
  484, 5734, \dodoi{10.1093/mnras/stz381}

\bibitem[{{Krumholz} \& {Kruijssen}(2015)}]{Krumholz:2015}
{Krumholz}, M.~R., \& {Kruijssen}, J.~M.~D. 2015, \mnras, 453, 739,
  \dodoi{10.1093/mnras/stv1670}

\bibitem[{{Lang} {et~al.}(2002){Lang}, {Goss}, \& {Morris}}]{Lang:2002}
{Lang}, C.~C., {Goss}, W.~M., \& {Morris}, M. 2002, \aj, 124, 2677,
  \dodoi{10.1086/344159}

\bibitem[{{Lang} {et~al.}(1999){Lang}, {Morris}, \& {Echevarria}}]{Lang:1999}
{Lang}, C.~C., {Morris}, M., \& {Echevarria}, L. 1999, \apj, 526, 727,
  \dodoi{10.1086/308012}

\bibitem[{{LaRosa} {et~al.}(2000){LaRosa}, {Kassim}, {Lazio}, \&
  {Hyman}}]{LaRosa:2000}
{LaRosa}, T.~N., {Kassim}, N.~E., {Lazio}, T. J.~W., \& {Hyman}, S.~D. 2000,
  \aj, 119, 207, \dodoi{10.1086/301168}

\bibitem[{{Law} {et~al.}(2008){Law}, {Yusef-Zadeh}, {Cotton}, \&
  {Maddalena}}]{Law:2008}
{Law}, C.~J., {Yusef-Zadeh}, F., {Cotton}, W.~D., \& {Maddalena}, R.~J. 2008,
  \apjs, 177, 255, \dodoi{10.1086/533587}

\bibitem[{{Li} {et~al.}(2018){Li}, {Austermann}, {Beall}, {Bruno}, {Choi},
  {Cothard}, {Crowley}, {Duff}, {Gallardo}, {Henderson}, {Ho}, {Hubmayr},
  {Koopman}, {McMahon}, {Niemack}, {Salatino}, {Simon}, {Staggs}, {Stevens},
  {Ullom}, {Ward}, \& {Wollack}}]{Li:2018}
{Li}, Y., {Austermann}, J.~E., {Beall}, J.~A., {et~al.} 2018, in Society of
  Photo-Optical Instrumentation Engineers (SPIE) Conference Series, Vol. 10708,
  \procspie, 107080A, \dodoi{10.1117/12.2313942}

\bibitem[{{Lis} {et~al.}(1994){Lis}, {Menten}, {Serabyn}, \&
  {Zylka}}]{Lis:1994b}
{Lis}, D.~C., {Menten}, K.~M., {Serabyn}, E., \& {Zylka}, R. 1994, \apjl, 423,
  L39, \dodoi{10.1086/187230}

\bibitem[{{Liszt}(2008)}]{Liszt:2008}
{Liszt}, H.~S. 2008, \aap, 486, 467, \dodoi{10.1051/0004-6361:200809748}

\bibitem[{{Liszt} \& {Spiker}(1995)}]{Liszt:1995}
{Liszt}, H.~S., \& {Spiker}, R.~W. 1995, \apjs, 98, 259, \dodoi{10.1086/192160}

\bibitem[{{Liszt} \& {Turner}(1978)}]{Liszt:1978}
{Liszt}, H.~S., \& {Turner}, B.~E. 1978, \apjl, 224, L73,
  \dodoi{10.1086/182762}

\bibitem[{{Longmore} {et~al.}(2012){Longmore}, {Rathborne}, {Bastian}, {Alves},
  {Ascenso}, {Bally}, {Testi}, {Longmore}, {Battersby}, {Bressert}, {Purcell},
  {Walsh}, {Jackson}, {Foster}, {Molinari}, {Meingast}, {Amorim}, {Lima},
  {Marques}, {Moitinho}, {Pinhao}, {Rebordao}, \& {Santos}}]{Longmore:2012}
{Longmore}, S.~N., {Rathborne}, J., {Bastian}, N., {et~al.} 2012, \apj, 746,
  117, \dodoi{10.1088/0004-637X/746/2/117}

\bibitem[{{Longmore} {et~al.}(2013){Longmore}, {Bally}, {Testi}, {Purcell},
  {Walsh}, {Bressert}, {Pestalozzi}, {Molinari}, {Ott}, {Cortese}, {Battersby},
  {Murray}, {Lee}, {Kruijssen}, {Schisano}, \& {Elia}}]{Longmore:2013}
{Longmore}, S.~N., {Bally}, J., {Testi}, L., {et~al.} 2013, \mnras, 429, 987,
  \dodoi{10.1093/mnras/sts376}

\bibitem[{{Lowe} {et~al.}(2020){Lowe}, {Coppi}, {Ade}, {Ashton}, {Austermann},
  {Beall}, {Clark}, {Cox}, {Devlin}, {Dicker}, {Dober}, {Fanfani}, {Fissel},
  {Galitzki}, {Gao}, {Hensley}, {Hubmayr}, {Li}, {Li}, {Lourie}, {Martin},
  {Mauskopf}, {Nati}, {Novak}, {Pisano}, {Romualdez}, {Sinclair}, {Soler},
  {Tucker}, {Vissers}, {Wheeler}, {Williams}, \& {Zannoni}}]{Lowe:2020}
{Lowe}, I., {Coppi}, G., {Ade}, P. A.~R., {et~al.} 2020, in Society of
  Photo-Optical Instrumentation Engineers (SPIE) Conference Series, Vol. 11445,
  Society of Photo-Optical Instrumentation Engineers (SPIE) Conference Series,
  114457A, \dodoi{10.1117/12.2576146}

\bibitem[{{Lu} {et~al.}(2021){Lu}, {Li}, {Ginsburg}, {Longmore}, {Kruijssen},
  {Walker}, {Feng}, {Zhang}, {Battersby}, {Pillai}, {Mills}, {Kauffmann},
  {Cheng}, \& {Inutsuka}}]{Lu:2021}
{Lu}, X., {Li}, S., {Ginsburg}, A., {et~al.} 2021, \apj, 909, 177,
  \dodoi{10.3847/1538-4357/abde3c}

\bibitem[{{Macquart} {et~al.}(2006){Macquart}, {Bower}, {Wright}, {Backer}, \&
  {Falcke}}]{Macquart:2006}
{Macquart}, J.-P., {Bower}, G.~C., {Wright}, M. C.~H., {Backer}, D.~C., \&
  {Falcke}, H. 2006, \apjl, 646, L111, \dodoi{10.1086/506932}

\bibitem[{{Mangilli} {et~al.}(2019){Mangilli}, {Aumont}, {Bernard}, {Buzzelli},
  {de Gasperis}, {Durrive}, {Ferriere}, {Fo{\"e}nard}, {Hughes}, {Lacourt},
  {Misawa}, {Montier}, {Mot}, {Ristorcelli}, {Roussel}, {Ade}, {Alina}, {de
  Bernardis}, {de Gouveia Dal Pino}, {Dubois}, {Engel}, {Guillet}, {Hargrave},
  {Laureijs}, {Longval}, {Maffei}, {Magalhaes}, {Marty}, {Masi}, {Montel},
  {Pajot}, {P{\'e}rot}, {Rodriguez}, {Salatino}, {Saccoccio}, {Savini},
  {Stever}, {Tauber}, {Tibbs}, \& {Tucker}}]{Mangilli:2019}
{Mangilli}, A., {Aumont}, J., {Bernard}, J.~P., {et~al.} 2019, \aap, 630, A74,
  \dodoi{10.1051/0004-6361/201935072}

\bibitem[{{Marrone} {et~al.}(2006){Marrone}, {Moran}, {Zhao}, \&
  {Rao}}]{Marrone:2006}
{Marrone}, D.~P., {Moran}, J.~M., {Zhao}, J.-H., \& {Rao}, R. 2006, in Journal
  of Physics Conference Series, Vol.~54, Journal of Physics Conference Series,
  354--362, \dodoi{10.1088/1742-6596/54/1/056}

\bibitem[{{Matthews} {et~al.}(2009){Matthews}, {McPhee}, {Fissel}, \&
  {Curran}}]{Matthews:2009}
{Matthews}, B.~C., {McPhee}, C.~A., {Fissel}, L.~M., \& {Curran}, R.~L. 2009,
  \apjs, 182, 143, \dodoi{10.1088/0067-0049/182/1/143}

\bibitem[{{Melia} {et~al.}(2000){Melia}, {Liu}, \& {Coker}}]{Melia:2000}
{Melia}, F., {Liu}, S., \& {Coker}, R. 2000, \apjl, 545, L117,
  \dodoi{10.1086/317888}

\bibitem[{{Melia} {et~al.}(2001){Melia}, {Liu}, \& {Coker}}]{Melia:2001}
---. 2001, \apj, 553, 146, \dodoi{10.1086/320644}

\bibitem[{{Miller} \& {LAMBDA}(2018)}]{Miller:2018}
{Miller}, N., \& {LAMBDA}. 2018, in American Astronomical Society Meeting
  Abstracts, Vol. 231, American Astronomical Society Meeting Abstracts \#231,
  430.05

\bibitem[{{Mills} {et~al.}(2015){Mills}, {Butterfield}, {Ludovici}, {Lang},
  {Ott}, {Morris}, \& {Schmitz}}]{Mills:2015}
{Mills}, E.~A.~C., {Butterfield}, N., {Ludovici}, D.~A., {et~al.} 2015, \apj,
  805, 72, \dodoi{10.1088/0004-637X/805/1/72}

\bibitem[{{Milne}(1970)}]{Milne:1970}
{Milne}, D.~K. 1970, Australian Journal of Physics, 23, 425,
  \dodoi{10.1071/PH700425}

\bibitem[{{Molinari} {et~al.}(2011){Molinari}, {Bally}, {Noriega-Crespo},
  {Compi{\`e}gne}, {Bernard}, {Paradis}, {Martin}, {Testi}, {Barlow}, {Moore},
  {Plume}, {Swinyard}, {Zavagno}, {Calzoletti}, {Di Giorgio}, {Elia},
  {Faustini}, {Natoli}, {Pestalozzi}, {Pezzuto}, {Piacentini}, {Polenta},
  {Polychroni}, {Schisano}, {Traficante}, {Veneziani}, {Battersby}, {Burton},
  {Carey}, {Fukui}, {Li}, {Lord}, {Morgan}, {Motte}, {Schuller},
  {Stringfellow}, {Tan}, {Thompson}, {Ward-Thompson}, {White}, \&
  {Umana}}]{Molinari:2011}
{Molinari}, S., {Bally}, J., {Noriega-Crespo}, A., {et~al.} 2011, \apjl, 735,
  L33, \dodoi{10.1088/2041-8205/735/2/L33}

\bibitem[{{Molinari} {et~al.}(2016){Molinari}, {Schisano}, {Elia},
  {Pestalozzi}, {Traficante}, {Pezzuto}, {Swinyard}, {Noriega-Crespo}, {Bally},
  {Moore}, {Plume}, {Zavagno}, {di Giorgio}, {Liu}, {Pilbratt}, {Mottram},
  {Russeil}, {Piazzo}, {Veneziani}, {Benedettini}, {Calzoletti}, {Faustini},
  {Natoli}, {Piacentini}, {Merello}, {Palmese}, {Del Grande}, {Polychroni},
  {Rygl}, {Polenta}, {Barlow}, {Bernard}, {Martin}, {Testi}, {Ali},
  {Andr{\'e}}, {Beltr{\'a}n}, {Billot}, {Carey}, {Cesaroni}, {Compi{\`e}gne},
  {Eden}, {Fukui}, {Garcia-Lario}, {Hoare}, {Huang}, {Joncas}, {Lim}, {Lord},
  {Martinavarro-Armengol}, {Motte}, {Paladini}, {Paradis}, {Peretto},
  {Robitaille}, {Schilke}, {Schneider}, {Schulz}, {Sibthorpe}, {Strafella},
  {Thompson}, {Umana}, {Ward-Thompson}, \& {Wyrowski}}]{Molinari:2016}
{Molinari}, S., {Schisano}, E., {Elia}, D., {et~al.} 2016, \aap, 591, A149,
  \dodoi{10.1051/0004-6361/201526380}

\bibitem[{{Morris} \& {Serabyn}(1996)}]{Morris:1996}
{Morris}, M., \& {Serabyn}, E. 1996, \araa, 34, 645,
  \dodoi{10.1146/annurev.astro.34.1.645}

\bibitem[{{Morris} \& {Yusef-Zadeh}(1985)}]{Morris:1985}
{Morris}, M., \& {Yusef-Zadeh}, F. 1985, \aj, 90, 2511, \dodoi{10.1086/113955}

\bibitem[{{Morris}(2015)}]{Morris:2015}
{Morris}, M.~R. 2015, {Manifestations of the Galactic Center Magnetic Field},
  391, \dodoi{10.1007/978-3-319-10614-4_32}

\bibitem[{Naess {et~al.}(2020)Naess, Aiola, Austermann, Battaglia, Beall,
  Becker, Bond, Calabrese, Choi, Cothard, \& et~al.}]{Naess:2020}
Naess, S., Aiola, S., Austermann, J.~E., {et~al.} 2020, Journal of Cosmology
  and Astroparticle Physics, 2020, 046–046,
  \dodoi{10.1088/1475-7516/2020/12/046}

\bibitem[{Nguyen {et~al.}(2021)Nguyen, Rugel, Menten, Brunthaler, Dzib, Yang,
  Kauffmann, Pillai, Nandakumar, Schultheis, Urquhart, Dokara, Gong, Medina,
  Ortiz-León, Reich, Wyrowski, Beuther, Cotton, Csengeri, Pandian, \&
  Roy}]{Nguyen:2021}
Nguyen, H., Rugel, M.~R., Menten, K.~M., {et~al.} 2021, A global view on star
  formation: The GLOSTAR Galactic plane survey IV. Radio continuum detections
  of young stellar objects in the Galactic Centre region.
\newblock \doarXiv{2105.03212}

\bibitem[{{Novak} {et~al.}(2000){Novak}, {Dotson}, {Dowell}, {Hildebrand},
  {Renbarger}, \& {Schleuning}}]{Novak:2000}
{Novak}, G., {Dotson}, J.~L., {Dowell}, C.~D., {et~al.} 2000, \apj, 529, 241,
  \dodoi{10.1086/308231}

\bibitem[{{Novak} {et~al.}(2003){Novak}, {Chuss}, {Renbarger}, {Griffin},
  {Newcomb}, {Peterson}, {Loewenstein}, {Pernic}, \& {Dotson}}]{Novak:2003}
{Novak}, G., {Chuss}, D.~T., {Renbarger}, T., {et~al.} 2003, \apjl, 583, L83,
  \dodoi{10.1086/368156}

\bibitem[{Næss(2019)}]{Naess:2019}
Næss, S.~K. 2019, Journal of Cosmology and Astroparticle Physics, 2019,
  060–060, \dodoi{10.1088/1475-7516/2019/12/060}

\bibitem[{{Oka} {et~al.}(2001){Oka}, {Hasegawa}, {Sato}, {Tsuboi}, \&
  {Miyazaki}}]{Oka:2001}
{Oka}, T., {Hasegawa}, T., {Sato}, F., {Tsuboi}, M., \& {Miyazaki}, A. 2001,
  \pasj, 53, 787, \dodoi{10.1093/pasj/53.5.787}

\bibitem[{{Par{\'e}} {et~al.}(2019){Par{\'e}}, {Lang}, {Morris}, {Moore}, \&
  {Mao}}]{Pare:2019}
{Par{\'e}}, D.~M., {Lang}, C.~C., {Morris}, M.~R., {Moore}, H., \& {Mao}, S.~A.
  2019, \apj, 884, 170, \dodoi{10.3847/1538-4357/ab45ed}

\bibitem[{{Parsons} {et~al.}(2018){Parsons}, {Dempsey}, {Thomas}, {Berry},
  {Currie}, {Friberg}, {Wouterloot}, {Chrysostomou}, {Graves}, {Tilanus},
  {Bell}, \& {Rawlings}}]{Parsons:2018}
{Parsons}, H., {Dempsey}, J.~T., {Thomas}, H.~S., {et~al.} 2018, \apjs, 234,
  22, \dodoi{10.3847/1538-4365/aa989c}

\bibitem[{{Pedlar} {et~al.}(1989){Pedlar}, {Anantharamaiah}, {Ekers}, {Goss},
  {van Gorkom}, {Schwarz}, \& {Zhao}}]{Pedlar:1989}
{Pedlar}, A., {Anantharamaiah}, K.~R., {Ekers}, R.~D., {et~al.} 1989, \apj,
  342, 769, \dodoi{10.1086/167635}

\bibitem[{{Pillai} {et~al.}(2015){Pillai}, {Kauffmann}, {Tan}, {Goldsmith},
  {Carey}, \& {Menten}}]{Pillai:2015}
{Pillai}, T., {Kauffmann}, J., {Tan}, J.~C., {et~al.} 2015, \apj, 799, 74,
  \dodoi{10.1088/0004-637X/799/1/74}

\bibitem[{{Planck Collaboration} {et~al.}(2020{\natexlab{a}}){Planck
  Collaboration}, {Aghanim}, {Akrami}, {Ashdown}, {Aumont}, {Baccigalupi},
  {Ballardini}, {Banday}, {Barreiro}, {Bartolo}, {Basak}, {Benabed}, {Bernard},
  {Bersanelli}, {Bielewicz}, {Bond}, {Borrill}, {Bouchet}, {Boulanger},
  {Bucher}, {Burigana}, {Calabrese}, {Cardoso}, {Carron}, {Challinor},
  {Chiang}, {Colombo}, {Combet}, {Couchot}, {Crill}, {Cuttaia}, {de Bernardis},
  {de Rosa}, {de Zotti}, {Delabrouille}, {Delouis}, {Di Valentino}, {Diego},
  {Dor{\'e}}, {Douspis}, {Ducout}, {Dupac}, {Efstathiou}, {Elsner},
  {En{\ss}lin}, {Eriksen}, {Falgarone}, {Fantaye}, {Finelli}, {Frailis},
  {Fraisse}, {Franceschi}, {Frolov}, {Galeotta}, {Galli}, {Ganga},
  {G{\'e}nova-Santos}, {Gerbino}, {Ghosh}, {Gonz{\'a}lez-Nuevo}, {G{\'o}rski},
  {Gratton}, {Gruppuso}, {Gudmundsson}, {Handley}, {Hansen},
  {Henrot-Versill{\'e}}, {Herranz}, {Hivon}, {Huang}, {Jaffe}, {Jones},
  {Karakci}, {Keih{\"a}nen}, {Keskitalo}, {Kiiveri}, {Kim}, {Kisner},
  {Krachmalnicoff}, {Kunz}, {Kurki-Suonio}, {Lagache}, {Lamarre}, {Lasenby},
  {Lattanzi}, {Lawrence}, {Levrier}, {Liguori}, {Lilje}, {Lindholm},
  {L{\'o}pez-Caniego}, {Ma}, {Mac{\'\i}as-P{\'e}rez}, {Maggio}, {Maino},
  {Mandolesi}, {Mangilli}, {Martin}, {Mart{\'\i}nez-Gonz{\'a}lez}, {Matarrese},
  {Mauri}, {McEwen}, {Melchiorri}, {Mennella}, {Migliaccio},
  {Miville-Desch{\^e}nes}, {Molinari}, {Moneti}, {Montier}, {Morgante}, {Moss},
  {Mottet}, {Natoli}, {Pagano}, {Paoletti}, {Partridge}, {Patanchon},
  {Patrizii}, {Perdereau}, {Perrotta}, {Pettorino}, {Piacentini}, {Puget},
  {Rachen}, {Reinecke}, {Remazeilles}, {Renzi}, {Rocha}, {Roudier}, {Salvati},
  {Sandri}, {Savelainen}, {Scott}, {Sirignano}, {Sirri}, {Spencer}, {Sunyaev},
  {Suur-Uski}, {Tauber}, {Tavagnacco}, {Tenti}, {Toffolatti}, {Tomasi},
  {Tristram}, {Trombetti}, {Valiviita}, {Vansyngel}, {Van Tent}, {Vibert},
  {Vielva}, {Villa}, {Vittorio}, {Wandelt}, {Wehus}, \&
  {Zonca}}]{Planck_l03:2020}
{Planck Collaboration}, {Aghanim}, N., {Akrami}, Y., {et~al.}
  2020{\natexlab{a}}, \aap, 641, A3, \dodoi{10.1051/0004-6361/201832909}

\bibitem[{{Planck Collaboration} {et~al.}(2020{\natexlab{b}}){Planck
  Collaboration}, {Aghanim}, {Akrami}, {Alves}, {Ashdown}, {Aumont},
  {Baccigalupi}, {Ballardini}, {Banday}, {Barreiro}, {Bartolo}, {Basak},
  {Benabed}, {Bernard}, {Bersanelli}, {Bielewicz}, {Bock}, {Bond}, {Borrill},
  {Bouchet}, {Boulanger}, {Bracco}, {Bucher}, {Burigana}, {Calabrese},
  {Cardoso}, {Carron}, {Chary}, {Chiang}, {Colombo}, {Combet}, {Crill},
  {Cuttaia}, {de Bernardis}, {de Zotti}, {Delabrouille}, {Delouis}, {Di
  Valentino}, {Dickinson}, {Diego}, {Dor{\'e}}, {Douspis}, {Ducout}, {Dupac},
  {Efstathiou}, {Elsner}, {En{\ss}lin}, {Eriksen}, {Falgarone}, {Fantaye},
  {Fernandez-Cobos}, {Ferri{\`e}re}, {Finelli}, {Forastieri}, {Frailis},
  {Fraisse}, {Franceschi}, {Frolov}, {Galeotta}, {Galli}, {Ganga},
  {G{\'e}nova-Santos}, {Gerbino}, {Ghosh}, {Gonz{\'a}lez-Nuevo}, {G{\'o}rski},
  {Gratton}, {Green}, {Gruppuso}, {Gudmundsson}, {Guillet}, {Handley},
  {Hansen}, {Helou}, {Herranz}, {Hivon}, {Huang}, {Jaffe}, {Jones},
  {Keih{\"a}nen}, {Keskitalo}, {Kiiveri}, {Kim}, {Krachmalnicoff}, {Kunz},
  {Kurki-Suonio}, {Lagache}, {Lamarre}, {Lasenby}, {Lattanzi}, {Lawrence}, {Le
  Jeune}, {Levrier}, {Liguori}, {Lilje}, {Lindholm}, {L{\'o}pez-Caniego},
  {Lubin}, {Ma}, {Mac{\'\i}as-P{\'e}rez}, {Maggio}, {Maino}, {Mandolesi},
  {Mangilli}, {Marcos-Caballero}, {Maris}, {Martin},
  {Mart{\'\i}nez-Gonz{\'a}lez}, {Matarrese}, {Mauri}, {McEwen}, {Melchiorri},
  {Mennella}, {Migliaccio}, {Miville-Desch{\^e}nes}, {Molinari}, {Moneti},
  {Montier}, {Morgante}, {Moss}, {Natoli}, {Pagano}, {Paoletti}, {Patanchon},
  {Perrotta}, {Pettorino}, {Piacentini}, {Polastri}, {Polenta}, {Puget},
  {Rachen}, {Reinecke}, {Remazeilles}, {Renzi}, {Ristorcelli}, {Rocha},
  {Rosset}, {Roudier}, {Rubi{\~n}o-Mart{\'\i}n}, {Ruiz-Granados}, {Salvati},
  {Sandri}, {Savelainen}, {Scott}, {Sirignano}, {Sunyaev}, {Suur-Uski},
  {Tauber}, {Tavagnacco}, {Tenti}, {Toffolatti}, {Tomasi}, {Trombetti},
  {Valiviita}, {Vansyngel}, {Van Tent}, {Vielva}, {Villa}, {Vittorio},
  {Wandelt}, {Wehus}, {Zacchei}, \& {Zonca}}]{Planck_l12:2020}
---. 2020{\natexlab{b}}, \aap, 641, A12, \dodoi{10.1051/0004-6361/201833885}

\bibitem[{{\sorthelp{Planck Collaboration 2014A}}{Planck Collaboration
  I}(2014)}]{planck2013-p01}
{\sorthelp{Planck Collaboration 2014A}}{Planck Collaboration I}. 2014, \aap,
  571, A1, \dodoi{10.1051/0004-6361/201321529}

\bibitem[{{\sorthelp{Planck Collaboration 2015J}}{Planck Collaboration
  X}(2016)}]{planck2014-a12}
{\sorthelp{Planck Collaboration 2015J}}{Planck Collaboration X}. 2016, \aap,
  594, A10, \dodoi{10.1051/0004-6361/201525967}

\bibitem[{{\sorthelp{Planck Collaboration IntS}}{Planck Collaboration Int.
  XIX}(2015)}]{planck2014-XIX}
{\sorthelp{Planck Collaboration IntS}}{Planck Collaboration Int. XIX}. 2015,
  \aap, 576, A104, \dodoi{10.1051/0004-6361/201424082}

\bibitem[{{\sorthelp{Planck Collaboration IntZZG}}{Planck Collaboration Int.
  LVII}(2020)}]{planck2020-LVII}
{\sorthelp{Planck Collaboration IntZZG}}{Planck Collaboration Int. LVII}. 2020,
  \aap, 643, 42, \dodoi{10.1051/0004-6361/202038073}

\bibitem[{{Plaszczynski} {et~al.}(2014){Plaszczynski}, {Montier}, {Levrier}, \&
  {Tristram}}]{Plaszczynski:2014}
{Plaszczynski}, S., {Montier}, L., {Levrier}, F., \& {Tristram}, M. 2014,
  \mnras, 439, 4048, \dodoi{10.1093/mnras/stu270}

\bibitem[{{Pound} \& {Yusef-Zadeh}(2018)}]{Pound:2018}
{Pound}, M.~W., \& {Yusef-Zadeh}, F. 2018, \mnras, 473, 2899,
  \dodoi{10.1093/mnras/stx2490}

\bibitem[{{Predehl} \& {Kulkarni}(1995)}]{Predehl:1995}
{Predehl}, P., \& {Kulkarni}, S.~R. 1995, \aap, 294, L29

\bibitem[{{Purcell}(1975)}]{Purcell:1975}
{Purcell}, E.~M. 1975, {Interstellar grains as pinwheels.}, ed. G.~B. {Field}
  \& A.~G.~W. {Cameron}, 155--167

\bibitem[{{Reich} {et~al.}(2000){Reich}, {Sofue}, \& {Matsuo}}]{Reich:2000}
{Reich}, W., {Sofue}, Y., \& {Matsuo}, H. 2000, \pasj, 52, 355,
  \dodoi{10.1093/pasj/52.2.355}

\bibitem[{{Roche} {et~al.}(2018){Roche}, {Lopez-Rodriguez}, {Telesco},
  {Sch{\"o}del}, \& {Packham}}]{Roche:2018}
{Roche}, P.~F., {Lopez-Rodriguez}, E., {Telesco}, C.~M., {Sch{\"o}del}, R., \&
  {Packham}, C. 2018, \mnras, 476, 235, \dodoi{10.1093/mnras/sty129}

\bibitem[{{Rodriguez-Fernandez} {et~al.}(2006){Rodriguez-Fernandez}, {Combes},
  {Martin-Pintado}, {Wilson}, \& {Apponi}}]{Rodriguez-Fernandez:2006}
{Rodriguez-Fernandez}, N.~J., {Combes}, F., {Martin-Pintado}, J., {Wilson},
  T.~L., \& {Apponi}, A. 2006, \aap, 455, 963,
  \dodoi{10.1051/0004-6361:20064813}

\bibitem[{{Ruud} {et~al.}(2015){Ruud}, {Fuskeland}, {Wehus}, {Vidal}, {Araujo},
  {Bischoff}, {Buder}, {Chinone}, {Cleary}, {Dumoulin}, {Kusaka}, {Monsalve},
  {N{\ae}ss}, {Newburgh}, {Reeves}, {Zwart}, {Bronfman}, {Davies}, {Davis},
  {Dickinson}, {Eriksen}, {Gaier}, {Gundersen}, {Hasegawa}, {Hazumi},
  {Huffenberger}, {Jones}, {Lawrence}, {Leitch}, {Limon}, {Miller}, {Pearson},
  {Piccirillo}, {Radford}, {Readhead}, {Samtleben}, {Seiffert}, {Shepherd},
  {Staggs}, {Tajima}, {Thompson}, \& {QUIET Collaboration}}]{Ruud:2015}
{Ruud}, T.~M., {Fuskeland}, U., {Wehus}, I.~K., {et~al.} 2015, \apj, 811, 89,
  \dodoi{10.1088/0004-637X/811/2/89}

\bibitem[{{Sandqvist}(1989)}]{Sandqvist:1989}
{Sandqvist}, A. 1989, \aap, 223, 293

\bibitem[{{Schuller} {et~al.}(2021){Schuller}, {Urquhart}, {Csengeri},
  {Colombo}, {Duarte-Cabral}, {Mattern}, {Ginsburg}, {Pettitt}, {Wyrowski},
  {Anderson}, {Azagra}, {Barnes}, {Beltran}, {Beuther}, {Billington},
  {Bronfman}, {Cesaroni}, {Dobbs}, {Eden}, {Lee}, {Medina}, {Menten}, {Moore},
  {Montenegro-Montes}, {Ragan}, {Rigby}, {Riener}, {Russeil}, {Schisano},
  {Sanchez-Monge}, {Traficante}, {Zavagno}, {Agurto}, {Bontemps}, {Finger},
  {Giannetti}, {Gonzalez}, {Hernandez}, {Henning}, {Kainulainen}, {Kauffmann},
  {Leurini}, {Lopez}, {Mac-Auliffe}, {Mazumdar}, {Molinari}, {Motte}, {Muller},
  {Nguyen-Luong}, {Parra}, {Perez-Beaupuits}, {Schilke}, {Schneider}, {Suri},
  {Testi}, {Torstensson}, {Veena}, {Venegas}, {Wang}, \&
  {Wienen}}]{Schuller:2021}
{Schuller}, F., {Urquhart}, J.~S., {Csengeri}, T., {et~al.} 2021, \mnras, 500,
  3064, \dodoi{10.1093/mnras/staa2369}

\bibitem[{{Scoville} {et~al.}(1975){Scoville}, {Solomon}, \&
  {Penzias}}]{Scoville:1975}
{Scoville}, N.~Z., {Solomon}, P.~M., \& {Penzias}, A.~A. 1975, \apj, 201, 352,
  \dodoi{10.1086/153892}

\bibitem[{{Serabyn} \& {Morris}(1994)}]{Serabyn:1994}
{Serabyn}, E., \& {Morris}, M. 1994, \apjl, 424, L91, \dodoi{10.1086/187282}

\bibitem[{{Shaver} {et~al.}(1985){Shaver}, {Salter}, {Patnaik}, {van Gorkom},
  \& {Hunt}}]{Shaver:1985}
{Shaver}, P.~A., {Salter}, C.~J., {Patnaik}, A.~R., {van Gorkom}, J.~H., \&
  {Hunt}, G.~C. 1985, \nat, 313, 113, \dodoi{10.1038/313113a0}

\bibitem[{{Simon} {et~al.}(2018){Simon}, {Beall}, {Cothard}, {Duff},
  {Gallardo}, {Ho}, {Hubmayr}, {Koopman}, {McMahon}, {Nati}, {Niemack},
  {Staggs}, {Vavagiakis}, \& {Wollack}}]{Simon:2018}
{Simon}, S.~M., {Beall}, J.~A., {Cothard}, N.~F., {et~al.} 2018, Journal of Low
  Temperature Physics, 193, 1041, \dodoi{10.1007/s10909-018-1963-7}

\bibitem[{{Sormani} \& {Barnes}(2019)}]{Sormani:2019}
{Sormani}, M.~C., \& {Barnes}, A.~T. 2019, \mnras, 484, 1213,
  \dodoi{10.1093/mnras/stz046}

\bibitem[{{Staguhn} {et~al.}(2019){Staguhn}, {Arendt}, {Dwek}, {Morris},
  {Yusef-Zadeh}, {Benford}, {Kov{\'a}cs}, \& {Gonzalez-Quiles}}]{Staguhn:2019}
{Staguhn}, J., {Arendt}, R.~G., {Dwek}, E., {et~al.} 2019, \apj, 885, 72,
  \dodoi{10.3847/1538-4357/ab451b}

\bibitem[{{Stevens} {et~al.}(2018){Stevens}, {Goeckner-Wald}, {Keskitalo},
  {McCallum}, {Ali}, {Borrill}, {Brown}, {Chinone}, {Gallardo}, {Kusaka},
  {Lee}, {McMahon}, {Niemack}, {Page}, {Puglisi}, {Salatino}, {Mak}, {Teply},
  {Thomas}, {Vavagiakis}, {Wollack}, {Xu}, \& {Zhu}}]{Stevens:2018}
{Stevens}, J.~R., {Goeckner-Wald}, N., {Keskitalo}, R., {et~al.} 2018, in
  Society of Photo-Optical Instrumentation Engineers (SPIE) Conference Series,
  Vol. 10708, Millimeter, Submillimeter, and Far-Infrared Detectors and
  Instrumentation for Astronomy IX, ed. J.~{Zmuidzinas} \& J.-R. {Gao},
  1070841, \dodoi{10.1117/12.2313898}

\bibitem[{{Stolovy} {et~al.}(1996){Stolovy}, {Hayward}, \&
  {Herter}}]{Stolovy:1996}
{Stolovy}, S.~R., {Hayward}, T.~L., \& {Herter}, T. 1996, \apjl, 470, L45,
  \dodoi{10.1086/310285}

\bibitem[{{Takekawa} {et~al.}(2014){Takekawa}, {Oka}, {Tanaka}, {Matsumura},
  {Miura}, \& {Sakai}}]{Takekawa:2014}
{Takekawa}, S., {Oka}, T., {Tanaka}, K., {et~al.} 2014, \apjs, 214, 2,
  \dodoi{10.1088/0067-0049/214/1/2}

\bibitem[{{Tanaka} {et~al.}(2007){Tanaka}, {Kamegai}, {Nagai}, \&
  {Oka}}]{Tanaka:2007}
{Tanaka}, K., {Kamegai}, K., {Nagai}, M., \& {Oka}, T. 2007, \pasj, 59, 323,
  \dodoi{10.1093/pasj/59.2.323}

\bibitem[{{Thornton} {et~al.}(2016){Thornton}, {Ade}, {Aiola}, {Angil{\`e}},
  {Amiri}, {Beall}, {Becker}, {Cho}, {Choi}, {Corlies}, {Coughlin}, {Datta},
  {Devlin}, {Dicker}, {D{\"u}nner}, {Fowler}, {Fox}, {Gallardo}, {Gao},
  {Grace}, {Halpern}, {Hasselfield}, {Henderson}, {Hilton}, {Hincks}, {Ho},
  {Hubmayr}, {Irwin}, {Klein}, {Koopman}, {Li}, {Louis}, {Lungu}, {Maurin},
  {McMahon}, {Munson}, {Naess}, {Nati}, {Newburgh}, {Nibarger}, {Niemack},
  {Niraula}, {Nolta}, {Page}, {Pappas}, {Schillaci}, {Schmitt}, {Sehgal},
  {Sievers}, {Simon}, {Staggs}, {Tucker}, {Uehara}, {van Lanen}, {Ward}, \&
  {Wollack}}]{Thornton:2016}
{Thornton}, R.~J., {Ade}, P.~A.~R., {Aiola}, S., {et~al.} 2016, \apjs, 227, 21,
  \dodoi{10.3847/1538-4365/227/2/21}

\bibitem[{{Tsujimoto} {et~al.}(2021){Tsujimoto}, {Oka}, {Takekawa}, {Iwata},
  {Uruno}, {Yokozuka}, {Nakagawara}, {Watanabe}, {Kawakami}, {Nishiyama},
  {Kaneko}, {Kanno}, \& {Ogawa}}]{Tsujimoto:2021}
{Tsujimoto}, S., {Oka}, T., {Takekawa}, S., {et~al.} 2021, \apj, 910, 61,
  \dodoi{10.3847/1538-4357/abe61e}

\bibitem[{{Uchida} {et~al.}(1992){Uchida}, {Morris}, \&
  {Yusef-Zadeh}}]{Uchida:1992}
{Uchida}, K., {Morris}, M., \& {Yusef-Zadeh}, F. 1992, \aj, 104, 1533,
  \dodoi{10.1086/116337}

\bibitem[{{Walker} {et~al.}(2021){Walker}, {Longmore}, {Bally}, {Ginsburg},
  {Kruijssen}, {Zhang}, {Henshaw}, {Lu}, {Alves}, {Barnes}, {Battersby},
  {Beuther}, {Contreras}, {G{\'o}mez}, {Ho}, {Jackson}, {Kauffmann}, {Mills},
  \& {Pillai}}]{Walker:2021}
{Walker}, D.~L., {Longmore}, S.~N., {Bally}, J., {et~al.} 2021, \mnras, 503,
  77, \dodoi{10.1093/mnras/stab415}

\bibitem[{{Weiler} \& {Sramek}(1988)}]{Weiler:1988}
{Weiler}, K.~W., \& {Sramek}, R.~A. 1988, \araa, 26, 295,
  \dodoi{10.1146/annurev.aa.26.090188.001455}

\bibitem[{{Yusef-Zadeh} \& {Bally}(1987)}]{Yusef-Zadeh:1987a}
{Yusef-Zadeh}, F., \& {Bally}, J. 1987, \nat, 330, 455,
  \dodoi{10.1038/330455a0}

\bibitem[{{Yusef-Zadeh} \& {Gaensler}(2005)}]{Yusef-Zadeh:2005}
{Yusef-Zadeh}, F., \& {Gaensler}, B.~M. 2005, Advances in Space Research, 35,
  1129, \dodoi{10.1016/j.asr.2005.03.003}

\bibitem[{{Yusef-Zadeh} \& {Morris}(1987{\natexlab{a}})}]{Yusef-Zadeh:1987b}
{Yusef-Zadeh}, F., \& {Morris}, M. 1987{\natexlab{a}}, \apj, 322, 721,
  \dodoi{10.1086/165767}

\bibitem[{{Yusef-Zadeh} \& {Morris}(1987{\natexlab{b}})}]{Yusef-Zadeh:1987c}
---. 1987{\natexlab{b}}, \apj, 320, 545, \dodoi{10.1086/165572}

\bibitem[{{Yusef-Zadeh} {et~al.}(1984){Yusef-Zadeh}, {Morris}, \&
  {Chance}}]{Yusef-Zadeh:1984}
{Yusef-Zadeh}, F., {Morris}, M., \& {Chance}, D. 1984, \nat, 310, 557,
  \dodoi{10.1038/310557a0}

\bibitem[{{Zhu} {et~al.}(2021){Zhu}, {Bhandarkar}, {Coppi}, {Kofman},
  {Orlowski-Scherer}, {Xu}, {Adachi}, {Ade}, {Aiola}, {Austermann}, {Bazarko},
  {Beall}, {Bhimani}, {Bond}, {Chesmore}, {Choi}, {Connors}, {Cothard},
  {Devlin}, {Dicker}, {Dober}, {Duell}, {Duff}, {D{\"u}nner}, {Fabbian},
  {Galitzki}, {Gallardo}, {Golec}, {Haridas}, {Harrington}, {Healy}, {Ho},
  {Huber}, {Hubmayr}, {Iuliano}, {Johnson}, {Keatin}, {Kiuchi}, {Koopman},
  {Lashner}, {Lee}, {Li}, {Limon}, {Link}, {Lucas}, {McCarrick}, {Moore},
  {Nati}, {Newburgh}, {Niemack}, {Pierpaoli}, {Randall}, {Perez Sarmiento},
  {Saunders}, {Seibert}, {Sierra}, {Sonka}, {Spisak}, {Sutariya}, {Tajima},
  {Teply}, {Thornton}, {Tsan}, {Tucker}, {Ullom}, {Vavagiakis}, {Vissers},
  {Walker}, {Westbrook}, {Wollack}, \& {Zannoni}}]{Zhu:2021}
{Zhu}, N., {Bhandarkar}, T., {Coppi}, G., {et~al.} 2021, arXiv e-prints,
  arXiv:2103.02747.
\newblock \doarXiv{2103.02747}

\end{thebibliography}
\bibliographystyle{aasjournal}

\end{document}